\documentclass[superscriptaddress,aps,preprint,amsmath,amssymb,floatfix]{revtex4}
\usepackage{graphicx,morefloats}
\usepackage{subfigure} % refering to Figure a and b separately
\usepackage{xcolor,soul}
\usepackage{caption}
\usepackage{bm}
\usepackage{helvet}
\usepackage{hyperref}
\definecolor{bg0}{rgb}{0,0,0}
\definecolor{bg}{rgb}{0,0,0.3}
\definecolor{bg1}{rgb}{1,1,1}
\definecolor{darkgreen}{rgb}{0,0.6,0.15}
\definecolor{mycolor}{rgb}{0.98, 0.0, 0.75}% Rule colour
\usepackage{caption,newfloat}
\DeclareCaptionType{InfoBox}

\begin{document}

\hyphenation{va-ni-sh-ing ana-ly-zing}

\begin{center}

\thispagestyle{empty}

{\large\bf Quantum phases driven by strong correlations}
\\[0.3cm]

Silke Paschen$^{1,2}$ and Qimiao Si$^{2}$\\[0.3cm]

$^1$Institute of Solid State Physics, Vienna University of Technology, 1040
Vienna, Austria\\

$^2$Department of Physics and Astronomy, Rice University, Houston, TX 77005,
USA\\

\end{center}

\vspace{0.0cm}

{\bf It has long been thought that strongly correlated systems are adiabatically
connected to their noninteracting counterpart. Recent developments have 
highlighted the fallacy of this traditional notion in a variety of settings.
Here we use a class of strongly correlated electron systems as a platform to
illustrate the kind of quantum phases and fluctuations that are created by
strong correlations. Examples are quantum critical states that violate the Fermi
liquid paradigm, unconventional superconductivity that goes beyond the BCS
framework, and topological semimetals induced by the Kondo interaction. We
assess the prospect of designing other exotic phases of matter, by utilizing
alternative degrees of freedom or alternative interactions, and point to the
potential of these correlated states for quantum technology.}
\vspace{0.4cm}

\noindent E-mails: paschen@ifp.tuwien.ac.at, qmsi@rice.edu 

%%%%%%%%%%%%%%%%%%%%%%%%%%%%%%%%%%%%%%%%%%%%%%%%%%%%%%%%%%%%%%%%%%%%%%%%%%%%%%
%%%%%%%%%%%% SECTION I %%%%%%%%%%%%%%%%%%%%%%%%%%%%%%%%%%%%%%%%%%%%%%%%%%%%%%%
%%%%%%%%%%%%%%%%%%%%%%%%%%%%%%%%%%%%%%%%%%%%%%%%%%%%%%%%%%%%%%%%%%%%%%%%%%%%%%

\section{Introduction}\label{introduction}

This review is motivated by a number of recent developments in the field of
strongly correlated electron systems that have created considerable excitement
in the community, and may stimulate progress also in adjacent fields. One of the
key open problems is the understanding of high-temperature superconductivity.
The recent discovery of a change in the carrier density at the pseudogap
critical point in a cuprate superconductor near optimal doping \cite{Bad16.1}
suggests a compelling link to the selective $4f$ Mott transition in heavy
fermion compounds across Kondo destruction quantum critical points (QCPs)
\cite{Sch00.1,Pas04.1,Shi05.1,Fri10.2,Cus12.1,Pro20.1}, which may also feature a
dome of high-temperature superconductivity \cite{Par06.1}. Similar phenomena are
observed in an organic superconductor \cite{Oik15.1} and, tentatively, in
twisted bilayer graphene \cite{Cao18.1}, further underpinning the need for a
unified understanding. Common principles are also surfacing with respect to
orbital physics. Being well established in transition metal oxides
\cite{Tok00.1}, orbitals are frequently entwined with other degrees of freedom
and may thus reveal their character only in the low-energy excitations. Examples
are the iron-based superconductors \cite{Si16.1}, the hidden order phase in
URu$_2$Si$_2$ \cite{Kun16.1}, or the sequential localization in an SU(4)-type
heavy fermion system \cite{Mar19.1}. Finally, there is a surge of work on the
interplay of strong correlations and topology. At focus so far have been quantum
spin liquids and topological insulators and semimetals in transition metal
compounds \cite{Sch16.2,Sav17.1,Zho17.1}, Weyl-Kondo semimetals in heavy fermion
systems \cite{Dzs17.1,Lai18.1}, and renormalized Dirac fermions in twisted
bilayer graphene \cite{Cao18.1}. It is expected that in the strong correlation
regime, a wealth of entirely new phases---not adiabatically connected to those
known in noninteracting systems---will be discovered, and push the frontiers of
science much like the discovery of the fractional quantum Hall effect
\cite{Sto99.1} did in 2-dimensional (2D) insulators.

So what is it that makes strongly correlated electron systems such a fertile
ground for discovery? Electrons are considered as strongly correlated when their
mutual Coulomb interaction is of the same order as their kinetic energy. This
does not only make the problem challenging---as simple one-electron theories
will fail---but rich. Tipping the balance between these two energies may change
the system's ground state as well as the nature of its low-energy excitations. A
particularly promising setting for emergent phenomena are continuous
zero-temperature phase transitions. They generate quantum critical fluctuations
which, in turn, may produce exotic excitations and novel phases. As suggested by
these considerations, strongly correlated electron systems have been providing a
rich setting for conceptual and theoretical advances. Examples include highly
entangled phases of matter \cite{Wen17.1} and QCPs that go beyond the Landau
framework \cite{Si01.1,Col01.1,Sen04.1}.

In Fig.\,\ref{materials} we introduce different materials classes hosting
strongly correlated electrons. Prominent representatives are the cuprate
high-$T_{\rm{c}}$ superconductors \cite{Kei15.2}, ruthenates and other
transition metal oxides \cite{Tok00.1}, heavy fermion systems \cite{Si10.2},
iron-based superconductors \cite{Si16.1}, organics \cite{Oik15.1}, and
low-dimensional materials \cite{Cao18.1}. The relevant electron orbitals in
these systems are typically $d$ or $f$ orbitals which, unlike $s$ or $p$
orbitals, keep some degree of localization in the solid and thus lead to
enhanced Coulomb interaction, and reduced band widths. In addition, at
appropriate filling, they give rise to localized moments. In
Sect.\,\ref{buildingblocks} we describe how these lead to the low-energy degrees
of freedom which---together with the conduction electrons, phonons, and other
excitations present also in weakly correlated cases---are the ``building
blocks'' of correlated matter. The mutual interactions between them, together
with the symmetry defined by the hosting crystal structure, result in an
overwhelming richness in physical properties. A key goal of the community is to
trace these properties back to the various magnetic, superconducting, and
topological phases (for representative phase diagrams, see Fig.\,S1), and
transitions between them, to ultimately allow to create functionality via
materials design. As we will discuss in Sect.\,\ref{tuningHF}, a viable approach
towards this goal is to apply non-thermal control parameters such as pressure or
field to a given material to tune it across its various phases, and to follow
the evolution of its properties. We will show that the extreme ``softness'' of
heavy fermion compounds to external parameters makes them particularly suitable
for such studies. The quantum phases and fluctuations that have been discovered
by such investigations are discussed in Sect.\,\ref{phases}. In
Sect.\,\ref{horizons} we describe new horizons in the field, addressing the
potential of alternative tuning parameters, additional degrees of freedom and
interactions beyond the canonical spin--conduction electron form, and the role
of topology. Implications of insight gained from studying heavy fermion systems
on other classes of correlated materials and, more generally, correlated states
of matter are discussed in Sect.\,\ref{implications}. We close with a summary
and outlook in Sect.\,\ref{summary}.

%%%%%%%%%%%%%%%%%%%%%%%%%%%%%%%%%%%%%%%%%%%%%%%%%%%%%%%%%%%%%%%%%%%%%%%%%%%%%%
%%%%%%%%%%%% SECTION II %%%%%%%%%%%%%%%%%%%%%%%%%%%%%%%%%%%%%%%%%%%%%%%%%%%%%%
%%%%%%%%%%%%%%%%%%%%%%%%%%%%%%%%%%%%%%%%%%%%%%%%%%%%%%%%%%%%%%%%%%%%%%%%%%%%%%

\section{Building blocks of correlated matter}\label{buildingblocks}

\begin{figure}[t!]
\centering

\includegraphics*[width=1\textwidth]{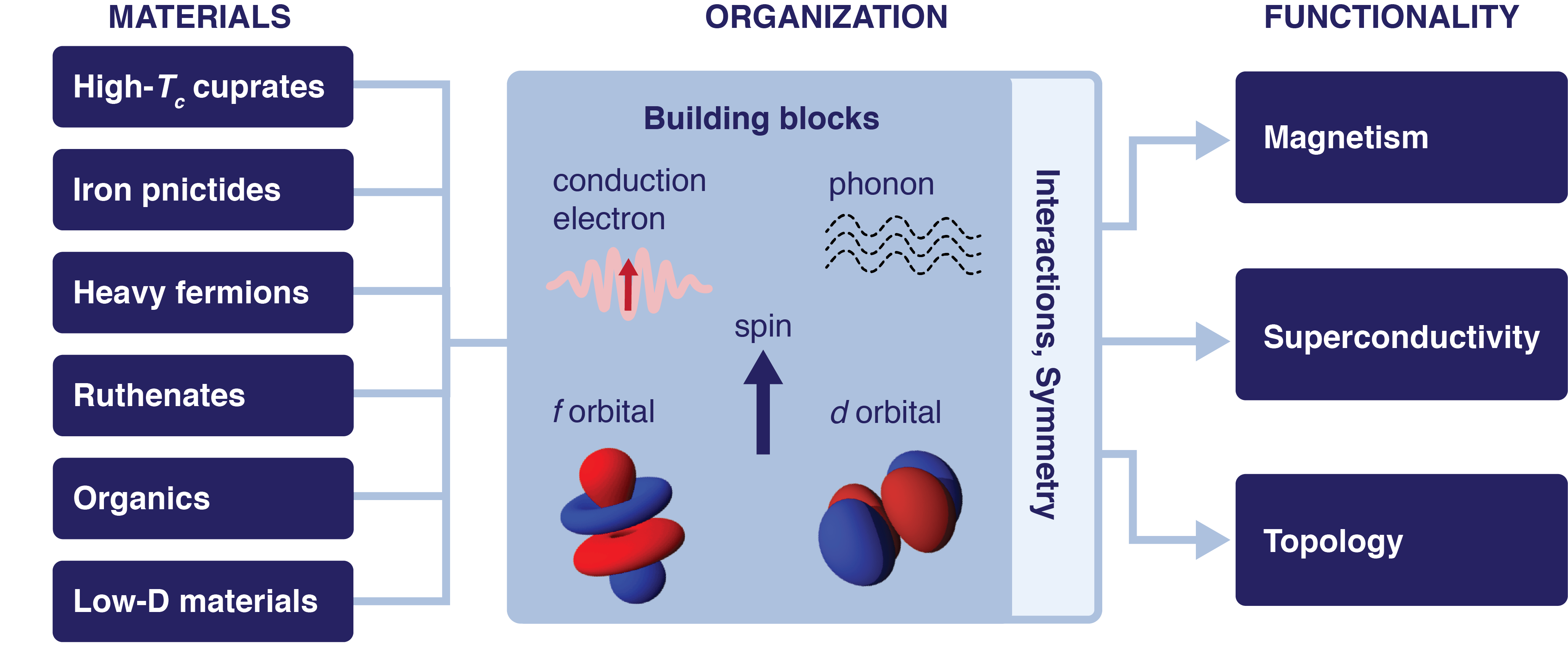}

\caption{\label{materials} {\bf Functionality of strongly correlated materials.}
Selected classes of strongly correlated materials, and how interactions between
their low-energy degrees of freedom (``building blocks'') and symmetry may lead
to different functionalities. Examples of the rich resulting phase diagrams for
several of these materials classes are shown in Fig.\,S1.}
\end{figure}

In strongly correlated electron systems, the ``onsite'' Coulomb
repulsion---between electrons associated with the same lattice site---is
comparable or larger than the width of the noninteracting energy band. Thus,
such atomic-level electron-electron interactions cannot be taken as a mere
perturbation. Instead, they are more appropriately treated at the zeroth order,
and a set of atomic levels ensues. These are energetically so far apart from
each other that only the lowest ones can directly influence the thermodynamic or
dynamical properties at low energies. The quantum numbers of these low-lying
atomic levels, together with those of delocalized excitations such as itinerant
electrons or acoustic phonons, specify the fundamental degrees of freedom---the
building blocks---of the system (Fig.\,\ref{materials}).

A familiar example is the Mott insulator in $3d$ transition metal systems. In
the simplest case, where only one $3d$ orbital is important for any given
lattice site, there is on average one electron per orbital (half-filled case),
and the onsite Coulomb repulsion $U$ is larger than the bare electron bandwidth.
The lowest-energy atomic level, a doublet with the degeneracy given by the spin
degree of freedom, is then occupied by a single electron; this corresponds to a
localized quantum spin $S=\frac{1}{2}$. Accessing the high-energy atomic levels
amounts to creating or removing an electron from the singly-occupied doublet.
Because this costs an energy on the order of $\frac{U}{2}$, these are
high-energy electronic excitations (responsible for the Hubbard bands) and do
not directly contribute to the low-energy properties. They do, however, mediate
superexchange interactions between the localized moments.

Another example arises in the $4f$ electron-based heavy fermion systems. The
strong onsite Coulomb repulsion of the $4f$ electrons, in combination with the
large spin-orbit coupling and crystal electric field (CEF) effects, frequently
lead to a Kramers doublet as the lowest-lying atomic level, again giving rise to
localized moments of effective spin $S=\frac{1}{2}$. These moments interact with
a separate band of $spd$ electrons, in the form of an antiferromagnetic Kondo
interaction, and with each other through a Ruderman-Kittel-Kasuya-Yosida (RKKY)
spin-exchange interaction that is mediated by the conduction electron spin
polarization \cite{Hew97.1}.

The presence of these well-defined building blocks and their interactions have
several implications. The Kondo and RKKY energy scales are small---much smaller
than the bare conduction electron bandwidth (Box\,\ref{box_Kondo} in
supplementary materials)---making the heavy fermion systems highly tunable and,
in turn, a prototype model setting for studying quantum criticality. In
addition, the clear understanding of the building blocks facilitates sharp
theoretical analyses. Thus, we use heavy fermion systems as a prototype setting
for our discussion of tuning and quantum criticality.

%%%%%%%%%%%%%%%%%%%%%%%%%%%%%%%%%%%%%%%%%%%%%%%%%%%%%%%%%%%%%%%%%%%%%%%%%%%%%%
%%%%%%%%%%%% SECTION III %%%%%%%%%%%%%%%%%%%%%%%%%%%%%%%%%%%%%%%%%%%%%%%%%%%%%
%%%%%%%%%%%%%%%%%%%%%%%%%%%%%%%%%%%%%%%%%%%%%%%%%%%%%%%%%%%%%%%%%%%%%%%%%%%%%%

\section{Tuning heavy fermion systems}\label{tuningHF}

There is a vast amount of correlated materials to be explored. This is due to
the large number of $d$ and $f$ elements in the periodic table which are---at
least in bulk materials---generally constituents of correlated materials, and
the numerous crystal structures of binary, ternary, quaternary (...) compounds
that may combine these elements with each other and with $s$ and $p$ elements.
Indeed, it is important to sample this huge ``phase space'' of correlated
materials at vastly different positions, to discover overarching principles and
test for universality. However, an alternative approach to this ``zoology'' has
proven particularly instructive: to (quasi-)continuously monitor the evolution
of a given system's properties along a specific path in the phase space. As will
be described in what follows such ``tuning'' studies have proven to be a
successful strategy to advance the field.

One goal of tuning studies is to establish the different phases or ground states
a given material system can adopt. A few examples of phase diagrams, displaying
the phase transition temperatures of the system as function of a non-thermal
tuning or control parameter, are shown in Fig.\,S1 of the supplementary
materials. This provides an overview and allows to relate the observed physical
properties to the different phases, or even to certain parts of a given phase
(e.g., center or border of the phase). Typical tuning parameters are chemical
substitutions within a family of isostructural compounds, doping as a special
case thereof, or the application of external parameters such as pressure or
magnetic field. Depending on the materials class, only modest or very large
variations of these parameters are needed to appreciably change the properties
of the system.

A materials class that is particularly responsive to such stimuli are heavy
fermion compounds. This is attributed to the presence of (at least) the
aforementioned two competing energy scales, the Kondo and the RKKY interaction
\cite{Hew97.1}. As already considered by Doniach for the simplified situation of
a one-dimensional Kondo lattice \cite{Don77.1}, tipping the balance between the
two scales is expected to tune the system from an antiferromagnetic to a
paramagnetic state. Indeed, numerous studies on heavy fermion metals have
identified such transitions
\cite{Ste01.1,Col05.1,Loe07.1,SpecialIssue2013,Si13.1,Kir20.1}.

Vital for establishing systematics, and thus for laying the ground for a
universal understanding, was the identification of a few simple, experimentally
well-defined, and meaningful characteristics of the studied materials. Since
early on \cite{And75.3} it is known that many heavy fermion metals show Fermi
liquid properties at low temperatures. The typically measured quantities are the
electrical resistivity, magnetic susceptibility, and specific heat. Their Fermi
liquid forms \cite{Hew97.1} are $\rho = \rho_0 + A T^2$, $\chi = \chi_0$, and $c
= \gamma T$. The correlation strength, quantified by the renormalized effective
mass $m^{\ast}$ of the charge carriers, enters in all three expressions, to
first approximation as $A \sim (m^{\ast})^2$, $\chi_0 \sim m^{\ast}$, and
$\gamma \sim m^{\ast}$. As such, the experimental determination of $A$,
$\chi_0$, and $\gamma$ allows to categorize heavy fermion metals according to
their degree of correlation. Assembling data of numerous heavy fermion metals in
$\gamma$ vs $\chi_0$ (Sommerfeld-Wilson) and $A$ vs $\gamma$ (Kadowaki-Woods)
plots confirmed the theoretically expected universal ratios \cite{Kad86.1}, in
particularly neat form if corrections due to different ground state degeneracies
\cite{Tsu05.1} and effects of dimensionality, electron density, and anisotropy
\cite{Jac09.1} are taken into account (Fig.\,\ref{tuning}a).

Further insight was gained with the discovery that the Fermi liquid parameters
of even a single given material can be widely varied by the application of
external tuning parameters. As illustrated with two examples in
Fig.\,\ref{tuning}b,c, in addition to ranges in tuning parameter where the
effective mass is continuously or ``adiabatically'' varied, there are
``neuralgic'' points where the effective mass appears to diverge. Clearly, at
such points, something extraordinary must be going on, and this is by now known
to be the presence of a QCP. Quantum critical fluctuations emerging from a QCP
lead to temperature dependencies distinct from Fermi liquid behavior, as will be
further discussed in Sect.\,\ref{phases}. The above-described divergence of the
Fermi liquid parameters on approaching a QCP is a valuable diagnostic tool
thereof. 

%%%%%%%%%%%%%%%%%%%%%%%%%%%%%%%%%%%%%%%%%%%%%%%%%%%%%%%%%%%%%%%%%%%%%%%%%%%%%%
%%%%%%%%%%%% SECTION IV %%%%%%%%%%%%%%%%%%%%%%%%%%%%%%%%%%%%%%%%%%%%%%%%%%%%%%%
%%%%%%%%%%%%%%%%%%%%%%%%%%%%%%%%%%%%%%%%%%%%%%%%%%%%%%%%%%%%%%%%%%%%%%%%%%%%%%

\section{Quantum phases and fluctuations}\label{phases}

A QCP appears as the ground state of a quantum many-body system smoothly
transforms from one state to another. Typical of the systems we consider in this
article is a gradual suppression of an electronic ordered phase
(Fig.\,\ref{entropy}a). On general grounds, entropy accumulates near a QCP. This
follows from hyperscaling: The free energy density, with the dimension of
energy/volume, is expected to have a singular component of the form
$\frac{k_{\rm B}T}{\xi^d}$ multiplied by a universal scaling function of
$\frac{|p-p_{\rm c}|}{\xi^{-1/\nu}}$ (here $\xi$ is the correlation length, $d$
the spacial dimension, $\nu$ the correlation length exponent and, for
concreteness, we consider the tuning parameter $\delta$ to be linearly coupled
to pressure $p$). It follows from this scaling form that the Gr\"uneisen ratio
as a function of $p$ is maximized near $p_{\rm c}$ \cite{Zhu03.1} and so is the
entropy $S$ (more precisely, the ratio $S/T$) for a sufficient low $T$,
Fig.\,\ref{entropy}b \cite{Wu11.1}.

The enhanced entropy makes the electron system particularly ``soft'', thereby
promoting the nucleation of novel excitations and phases. For instance, strongly
correlated metals display an unusual temperature dependence of the electrical
resistivity at the QCP, reflecting the breakdown of the Landau quasiparticle
concept and the ensuing formation of a non-Fermi liquid (``strange metal''). In
addition, unconventional superconductivity often develops near a QCP. As such,
quantum criticality rises to the level of a unifying theme for the description
of the often exotic properties across a variety of strongly correlated materials
\cite{SpecialIssue2013,Kei17.1,Bal17.1,Kir20.1}.

In practice, quantum criticality often develops in bad metals and in the
vicinity of antiferromagnetic order. A bad metal is specified by its resistivity
at room temperature being large, reaching the Mott-Ioffe-Regel limit, which
implicates the presence of strong electron correlations \cite{Hus04.1}. Such
correlations cause a reduction in the coherent part of the electron spectral
weight, as measured by angle resolved photoemission spectroscopy (ARPES)
\cite{Yi17.1} and optical conductivity \cite{Dre02.1,Qaz09.1}. Because a
sufficiently large correlation strength suppresses the coherent electron part,
it may lead to an electronic localization--delocalization transition. All these
phenomena are exemplified by heavy fermion metals, which are canonical examples
of bad metals and in which antiferromagnetic QCPs are especially prevalent
\cite{Ste01.1,Col05.1,Loe07.1, Si10.2,SpecialIssue2013,Si13.1, Kir20.1}. We thus
discuss these systems in what follows.

\subsection{Quantum criticality}

Consider a Kondo lattice (see Box\,\ref{box_Kondo} of supplementary materials),
comprising a localized moment for each unit cell and a band of conduction
electrons (Fig.\,\ref{FSjump}a,b). When the RKKY interaction dominates over the
Kondo coupling, the local moments develop static order and form an
antiferromagnetic state (Fig.\,\ref{FSjump}a). Deep in this state, it has been
rigorously shown that the Kondo coupling is ``exactly marginal'' in the
renormalization-group sense \cite{Yam07.1}; here, spin singlets are well
established among the local moments, which are detrimental to the development of
a Kondo singlet between the local moments and conduction electrons. The Fermi
surface is then formed of the conduction electrons only---this is named a
``small'' Fermi surface (Fig.\,\ref{FSjump}c).

In the opposite limit, when the Kondo coupling dominates over the RKKY
interaction, the local moments form a spin singlet state with the conduction
electrons (Fig.\,\ref{FSjump}b). This is the standard paramagnetic heavy fermion
state, characterized by a static Kondo-singlet amplitude. The ensuing Kondo
entanglement of the local moments with the conduction electrons effectively
turns the former into mobile carriers: A local moment and a conduction electron
form a fermionic composite, which corresponds to a delocalized heavy electronic
excitation. Its hybridization with the conduction electron band
(Fig.\,\ref{FSjump}f-h) produces the heavy quasiparticles. Thus, in this limit,
the Fermi surface incorporates both the local moments and conduction electrons,
and is ``large'' (Fig.\,\ref{FSjump}d).

What is remarkable is that the transformation from the small to the large Fermi
surface can occur sharply, precisely at a continuous zero-temperature phase
transition from an antiferromagnetically ordered to a paramagnetic heavy fermion
state. In YbRh$_2$Si$_2$, an antiferromagnet with a low N\'eel temperature
($T_{\rm N} = 0.07$ K), application of a small magnetic field weakens the
antiferromagnetic order and continuously suppresses it at the critical field
$B_{\rm c}$ (Fig.\,\ref{tuning}b). Hall effect measurements
\cite{Pas04.1,Fri10.2} imply a sudden jump in the extrapolated zero-temperature
limit, precisely at the QCP (Fig.\,\ref{FSjump}e), consistent with a sudden
change from the small to the large Fermi surface. At nonzero temperatures, the
Hall results as well as thermodynamic measurements \cite{Geg07.1} identify a
temperature scale $T^*(B)$ in the temperature--magnetic field phase diagram
which, anchored by the Fermi-surface-jumping QCP, characterizes the
finite-temperature crossovers. Scanning tunneling microscopy (STM) experiments
have provided spectroscopic evidence for the $T^*(B)$ scale \cite{Sei18.1},
where a critical slowing down was evidenced from a linewidth analysis.

Another experimental signature of this drastic effect was revealed by quantum
oscillation measurements. CeRhIn$_5$ is an antiferromagnetic metal with a N\'eel
temperature $T_{\rm N} = 3.8$\,K \cite{Heg00.1}. Applying pressure weakens the
antiferromagnetic order and induces superconductivity \cite{Par06.1,Kne08.1}. In
magnetic fields above the superconducting upper critical field $H_{\rm c2}$
these studies \cite{Par06.1,Kne08.1} reveal an antiferromagnetic QCP near the
same pressure $p_{\rm c}$ where the de Haas--van Alphen (dHvA) frequencies jump
and the cyclotron mass diverges \cite{Shi05.1}. From a comparison with {\it ab
initio} electronic structure calculations for LaRhIn$_5$ and CeRhIn$_5$ (with
the Ce-$4f$ electrons in the core) it was concluded that the Fermi surface
transforms from small at $p<p_{\rm c}$ to large at $p>p_{\rm c}$ \cite{Shi05.1}.

These striking experimental observations were anticipated by theoretical studies
\cite{Si01.1,Col01.1,Sen04.1}. Analyzing the Kondo lattice model identified an
unusual type of QCP, where a small-to-large Fermi surface jump occurs at the
antiferromagnetic-paramagnetic phase boundary \cite{Si01.1}. The quantum
criticality then captures not only the fluctuations from a suppression of the 
antiferromagnetic order parameter, but also those associated with a destruction
of the static Kondo effect.

\begin{figure}[h!]
\centerline{\hspace{-0.5cm}\includegraphics[height=0.36\textwidth]{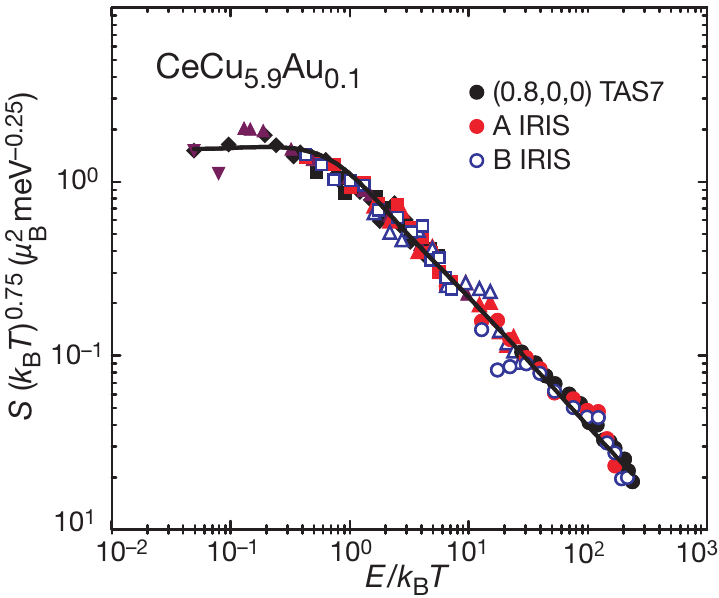}\hspace{0.7cm}\includegraphics[height=0.368\textwidth]{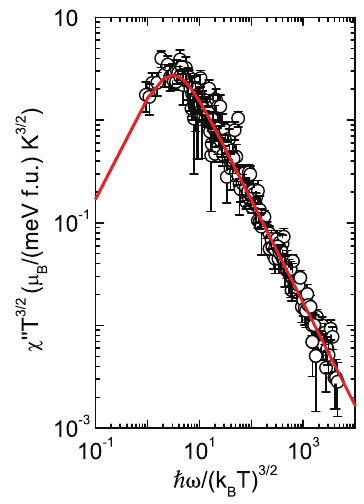}}
\vspace{0.5cm}

\centerline{\includegraphics[height=0.358\textwidth]{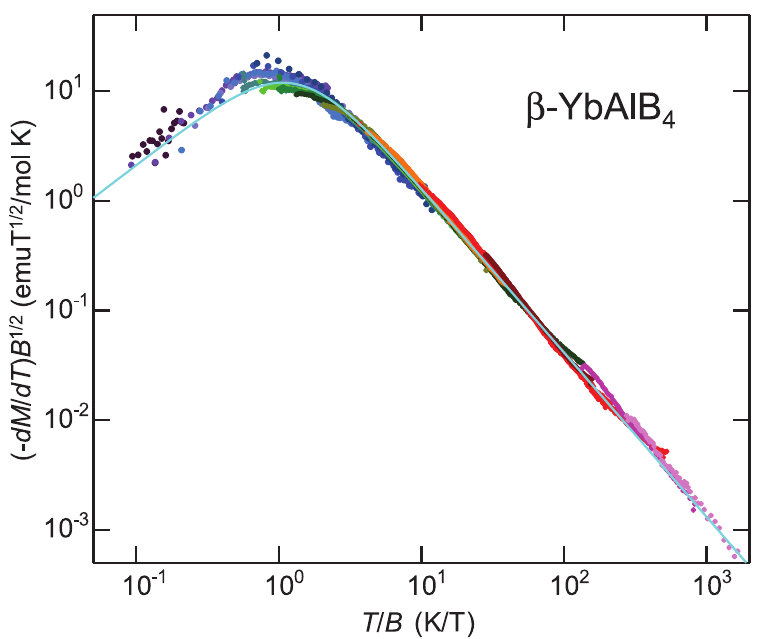}\hspace{0.5cm}\includegraphics[height=0.36\textwidth]{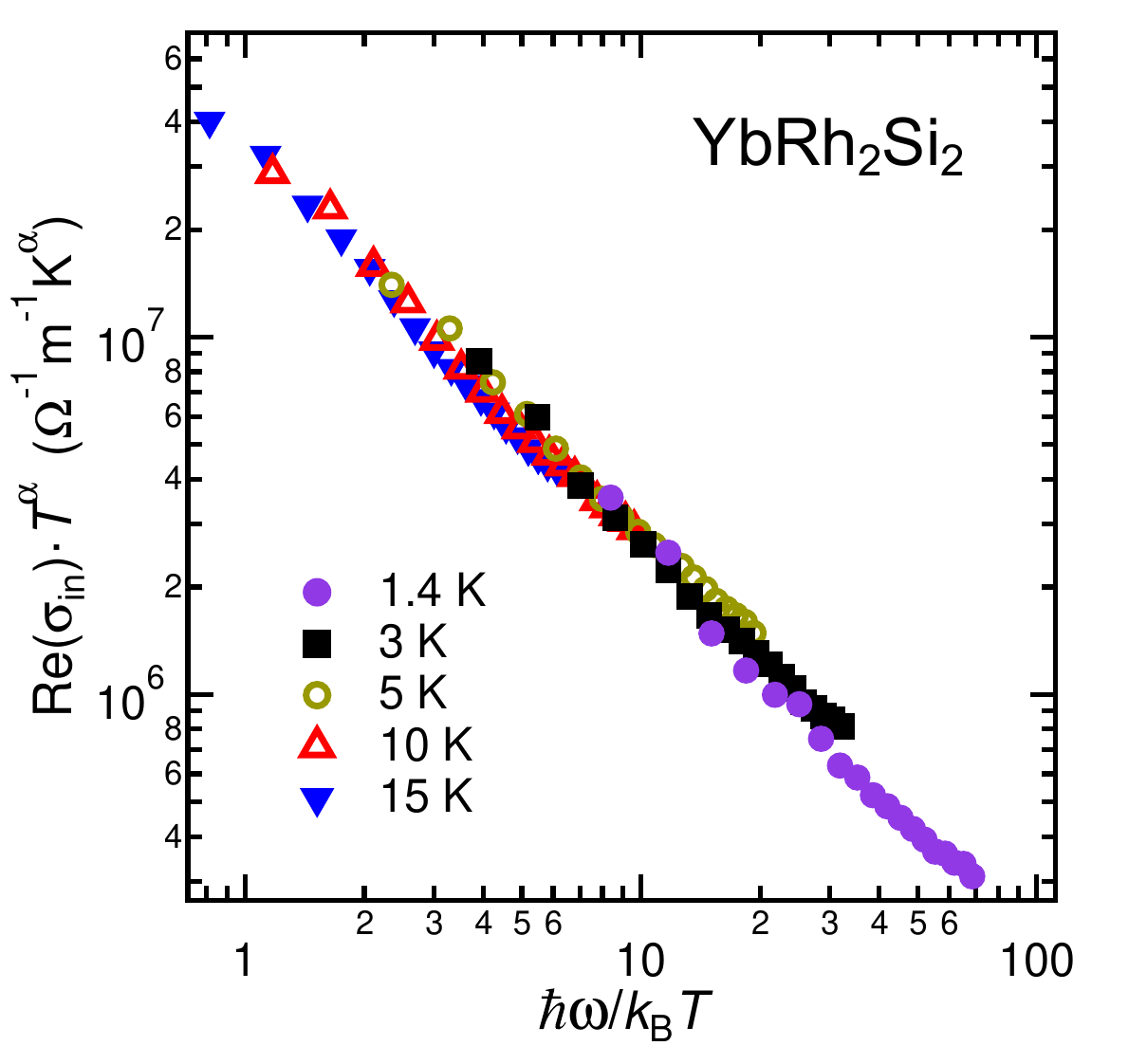}}
\vspace{-12.9cm}

\hspace{-4.8cm}{\bf\textsf{a}}\hspace{7.8cm}{\bf\textsf{b}}
\vspace{5.8cm}

\hspace{-6.3cm}{\bf\textsf{c}}\hspace{7.5cm}{\bf\textsf{d}}
\vspace{6.2cm}

\caption{\label{scaling} {\bf Quantum critical scaling.} {\bf a-b}$|$ Dynamical
scaling in the spin dynamics of CeCu$_{5.9}$Au$_{0.1}$ (adapted with permission
from ref.\citenum{Sch00.1}) and CeCu$_2$Si$_2$ (adapted with permission from
ref.\citenum{Arn11}). {\bf c}$|$ $T/B$ scaling of $d M/dT$, where $M$ is the
magnetization, in $\beta$-YbAlB$_4$ (adapted with permission from
ref.\citenum{Mat11.1}). {\bf d}$|$ $\omega/T$ scaling in the optical
conductivity of YbRh$_2$Si$_2$ (adapted with permission from
ref.\citenum{Pro20.1}).}
\end{figure}

Kondo destruction quantum criticality has another salient feature, an $\omega/T$
scaling in the spin dynamics. The collapse of the static Kondo singlet makes the
QCP to be described by an interacting fixed point, where the universal physics
depends on only $k_{\rm B}T$ and no other energy scales \cite{Si01.1,Col01.1}.
Calculations on Kondo lattice models based on an extended dynamical mean-field
theory (EDMFT) have produced such $\omega/T$ behavior in the dynamical spin
susceptibility at the Kondo destruction QCP \cite{Si01.1,Gre03.2,Hu20.1x}. This
provides the understanding of inelastic neutron scattering experiments on
quantum critical CeCu$_{5.9}$Au$_{0.1}$ (Fig.\,\ref{scaling}a). Its parent
compound CeCu$_6$ is a paramagnetic heavy fermion metal, and a substitution of
less than 2\% of Cu by Au induces antiferromagnetic order. In
CeCu$_{5.9}$Au$_{0.1}$, the dynamical spin susceptibility displays $\omega/T$
scaling with a fractional critical exponent (Fig.\,\ref{scaling}a). In
$\beta$-YbAlB$_4$, $B/T$ scaling has been shown in thermodynamic quantities
\cite{Mat11.1} (Fig.\,\ref{scaling}c), but dynamical scaling could not yet be
tested.

An exciting recent development came from terahertz spectroscopy measurements of
the optical conductivity of YbRh$_2$Si$_2$, which demonstrate $\omega/T$ scaling
in the charge channel (Fig.\,\ref{scaling}d) \cite{Pro20.1}. This was understood
as a consequence of probing a Kondo destruction QCP: Because the Kondo effect
involves also the charge degree of freedom, its destruction should also lead to
$\omega/T$ scaling in the charge sector.

The above features are to be contrasted with expectations for quantum
criticality within the Landau framework, which is of spin-density-wave (SDW)
type \cite{Her76.1,Mil93.1,Mor12.2}. Going across an SDW QCP from the
paramagnetic side, the Fermi surface evolves smoothly: it folds by the
continuously onsetting SDW order parameter. As a result, across this QCP, both
the Hall coefficient and the quantum oscillation frequencies vary smoothly.
Moreover, the SDW QCP is described by a Gaussian fixed point, and a dangerously
irrelevant variable generates an effective energy scale other than $k_{\rm B}T$.
In turn, $\omega/T$ scaling is violated.

Following the tradition of the paramagnon theory \cite{Lev83.1}, the SDW-type
QCP is discussed for purely itinerant electron systems. In heavy fermion
compounds, local moments are part of the building blocks for the low-energy
physics. It has been stressed that, for the Kondo destruction QCP to arise, it
is essential to treat the dynamical competition between the RKKY and Kondo
interactions. For instance, while the static Kondo singlet amplitude vanishes as
the Kondo destruction QCP is approached from the paramagnetic side, the Kondo
correlations at nonzero frequencies smoothly evolve across the QCP
\cite{Cai19.1x,Hu20.1x}. This effect is essential for the stability of the
Kondo-destroyed phase, and also provides the understanding of a mass enhancement
in that phase.

In view of this dynamical competition between the RKKY and Kondo interactions it
follows that even in heavy fermion metals described by an SDW QCP at the lowest
energies, the Kondo destruction crossover scale $T^*$, though nonzero, can still
be considerably smaller than the bare Kondo temperature scale $T_K^0$
\cite{Ney17.1}. In this case, there is an extended temperature range ($T^* < T <
T_K^0$) over which Kondo destruction quantum criticality should dominate the
singular physics. For instance, in CeCu$_2$Si$_2$, the dynamical spin
susceptibility follows the $\omega/T^{3/2}$ scaling form predicted for a
three-dimensional SDW QCP (Fig.\,\ref{scaling}b) \cite{Arn11}, corresponding to
a $T^{3/2}$ relaxation rate. However, this rate becomes linear in $T$ already at
1\,K and above, pointing to $T^* \sim 1$\,K and a sizable temperature range,
$T^*<T<T_K^0\sim 20$\,K where the dynamics may be described by Kondo destruction
physics; additional ways to ascertain this possibility are needed.

These developments establish antiferromagnetic heavy fermion metals to be a
prototype setting for quantum criticality beyond the Landau framework. The fact
that such a form of unconventional quantum criticality takes place at a
transition between two conventional phases has motivated studies of
beyond-Landau QCPs in the context of quantum magnets \cite{Sen04.2}. From the
correlated electron perspective, Kondo destruction signifies an electronic
localization--delocalization transition; indeed, it has been shown that the
Kondo destruction QCP features criticality in both the spin and the charge (or
single-particle) channels \cite{Si01.1,Gre03.2,Zhu04.1,Kom19.1,Cai20.2,Hu20.1x}.
We will discuss the significance of this physics for systems beyond heavy
fermion metals in Sec.\,\ref{implications}A. More broadly, the Kondo destruction
QCP is also captured by the evolution of entanglement entropy
\cite{Pix15.2,Wag18.1,Hu20.1x}; exploring entanglement properties, perhaps using
emulating set-ups based on cold atoms or mesoscopic devices, will deepen our
understanding of this novel type of quantum criticality and its associated
strong correlation physics.

\subsection{Unconventional superconductivity and other emergent phases}

As we have highlighted, quantum criticality provides a mechanism to soften the
electronic system and nucleate novel phases. A widely recognized case in point
is the development of unconventional superconductivity \cite{Mat98.1,Ste16.1}.
In CePd$_2$Si$_2$, a dome of superconductivity is observed around a
pressure-induced QCP \cite{Mat98.1}. This QCP was discussed as being of SDW type
and the superconducting pairing was associated with antiferromagnetic
paramagnons \cite{Sca12.1}. This led to the important question whether a Kondo
destruction QCP can also promote superconductivity. Theoretical calculations
suggest that this is indeed the case \cite{Pix15.3,Cai20.1}.

There are by now about fifty heavy fermion superconductors. Many of them seem to
be driven by antiferromagnetic quantum criticality, but whether that is indeed
the case and, if so, what is the nature of the underlying QCP, remains to be
clarified in most cases. In CeRhIn$_5$, where a Kondo destruction QCP was
evidenced at a critical pressure from (above-$H_{c2}$) dHvA experiments 
\cite{Shi05.1}, superconductivity indeed appeared near that pressure
\cite{Par06.1,Kne08.1}, suggesting that it is driven by the Kondo destruction
QCP. The optimal $T_{\rm c}$ is about 2.3\,K (in zero field)
\cite{Par06.1,Kne08.1}, which is among the highest transition temperatures for
$4f$ electron-based heavy fermion superconductors. 

CeCu$_2$Si$_2$ is the heavy fermion metal in which unconventional
superconductivity was first discovered \cite{Ste79.1}, with $T_{\rm c} =
0.6$\,K. The magnetic fluctuation spectrum changes noticeably upon entering the
superconducting state \cite{Sto11.1}, an effect that was also studied in other
heavy fermion superconductors \cite{Sto08.1}; this gives rise to a large gain in
the magnetic exchange energy, which is about $20$ times the superconducting
condensation energy \cite{Sto11.1,Sto12.4}, thereby providing evidence that the
quantum critical magnetic fluctuations are a major driving force for the
development of unconventional superconductivity. It also implies a loss of
kinetic energy of close to 20 times the superconducting condensation energy,
which is compatible with pairing in the spin-singlet channel. In that case, the
development of superconducting pairing weakens the Kondo-singlet correlation
and, thus, transfers the $f$-electron spectral weight from low energies to above
the Kondo-destruction energy scale $T^*$ \cite{Sto12.4}. The precise form of the
spin-singlet pairing order parameter has been reanalysed in recent years. A
multi-orbital $d+d$ pairing \cite{Nic17.1,Nic19.1x} provides a natural
understanding of the gap-like features that have recently been observed by
measurements of the specific heat and superfluid stiffness at the lowest
temperatures \cite{Pan18.1,Yam17.1,Kit14.1}. This pairing state obeys the
prohibition of any onsite pairing amplitude by the Coulomb repulsion. It
features a spin resonance in the superconducting state \cite{Sto11.1}, and the
correlation effect is expected to reduce its sensitivity to non-magnetic
disorder \cite{Smi18.1}. The conventional $s$-wave pairing, which has also been
invoked to explain the observed gap \cite{Yam17.1}, faces a large energetic
penalty. Extended $s$-wave pairings have also been considered
\cite{Ike15.2,Li18.2}. For the Fermi surface of CeCu$_2$Si$_2$, they are
expected to be nodal, with any spin resonance being away from the intraband
nesting wavevector observed experimentally \cite{Smi18.1}. Microscopically, the
$d+d$ pairing order parameter, an irreducible representation of the crystalline
point group, comprises intraband and interband $d$-wave pairing  components. It
is the non-commuting nature of the two components that produces a full gap on
the Fermi surface, which is analogous to what happens in the $^3$He superfluid
$B$-phase \cite{Nic19.1x}.

For a long time, no superconductivity was found in YbRh$_2$Si$_2$. This changed
when the specific heat and magnetization were measured at ultralow temperatures
\cite{Sch16.1}. An analysis of the coupling between nuclear and electronic spins
led to the picture that, below 2\,mK, superconductivity coexists with hybrid
nuclear--electron order, which reduces the primary electronic order and,
consequently, enhances quantum critical fluctuations.

Other forms of emergent phases may also develop near a QCP. In CeRhIn$_5$ at
ambient pressure, quantum oscillation and Hall effect measurements provided
evidence for a Kondo destruction QCP near $B_0^* \approx 30$\,T \cite{Jia15.1},
which is considerably below the field for the zero-temperature magnetic to
paramagnetic transition, $B_{c0} \approx 50$\,T. Near $B_0^*$, transport
measurements in samples fabricated by focused ion beam etching have indicated
the development of nematic correlations \cite{Ron17.1}. Whether static nematic
order exists and, if so, whether it is nucleated by Kondo destruction quantum
criticality are interesting questions that deserve further investigations.

%%%%%%%%%%%%%%%%%%%%%%%%%%%%%%%%%%%%%%%%%%%%%%%%%%%%%%%%%%%%%%%%%%%%%%%%%%%%%%
%%%%%%%%%%%% SECTION V %%%%%%%%%%%%%%%%%%%%%%%%%%%%%%%%%%%%%%%%%%%%%%%%%%%%%%%
%%%%%%%%%%%%%%%%%%%%%%%%%%%%%%%%%%%%%%%%%%%%%%%%%%%%%%%%%%%%%%%%%%%%%%%%%%%%%%

\section{New horizons}\label{horizons}

Many of the extensive developments on heavy fermion quantum criticality,
reviewed above, correspond to the prototypical setting \cite{Don77.1}, i.e.\ the
physics of the competition between Kondo and RKKY interaction in a (spin 1/2)
Kondo lattice. It is becoming increasingly clear that some of the most
interesting phenomena, such as quantum criticality associated with electron
localization and the formation of novel phases, are not limited to this
canonical case but can be generalized to much broader contexts. Some of these
developments will be highlighted in this section.

\subsection{Frustration and dimensionality}

\begin{figure}[b!]
\centering
\includegraphics[width=0.9\textwidth,]{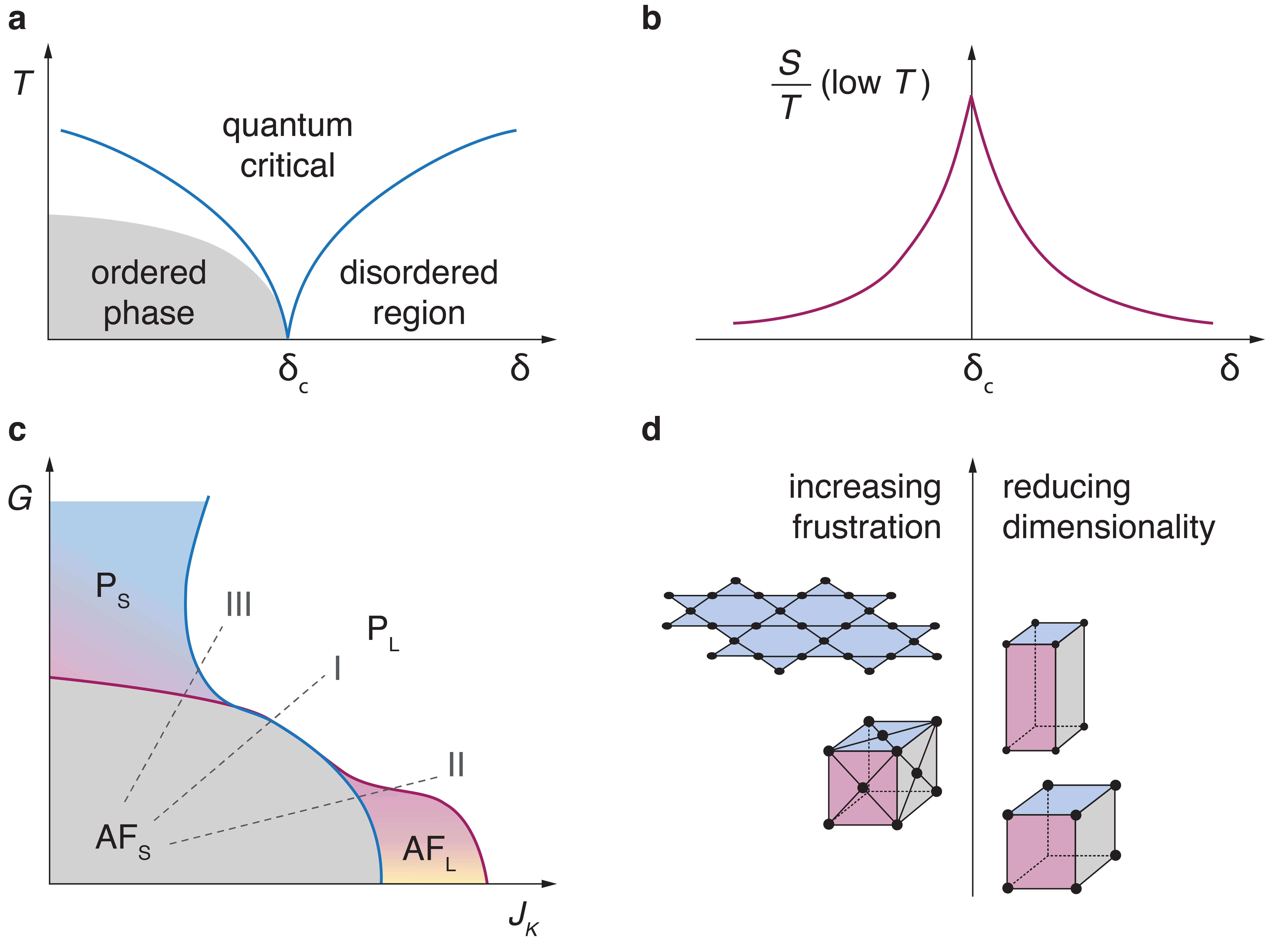}
\caption{\label{entropy} {\bf Quantum criticality.} {\bf a}$|$ Emergence of a
QCP in the temperature--tuning parameter ($T$ vs $\delta$) phase diagram. The
QCP, at zero temperature and $\delta=\delta_c$, represents the point where
long-range order is continuously suppressed to zero. {\bf b}$|$ Evolution of the
entropy $S$ vs $\delta$, showing that $S/T$ at a low $T$ is peaked near the QCP
(adapted with permission from ref.\citenum{Wu11.1}). {\bf c}$|$ A global phase
diagram for heavy fermion metals, where $J_{\rm K}$ tunes the Kondo coupling and
$G$ varies the degree of frustration. P and AF stand for paramagnetic and
antiferromagnetic phases, while the subscripts L and S denote the Fermi surface
being large and small (adapted with permission from refs.\citenum{Si06.1} and
\citenum{Si10.1}). {\bf d}$|$ Illustration of variations in the degree of
geometrical frustration and in spatial dimensionality (adapted with permission
from ref.\citenum{Cus12.1}).}
\end{figure}

The localization--delocalization transition discussed earlier originates from
the different types of correlations that are promoted by the Kondo and RKKY
interactions. The singlet formation among the spins in the antiferromagnetic
state leads to a destruction of the static Kondo effect. When the degree of
frustration $G$ \cite{Nis13.1} is large, spin singlets among the local moments 
develop even in a quantum paramagnet \cite{Sch16.2,Sav17.1,Zho17.1} and will be
detrimental to the formation of the Kondo entanglement. This line of
consideration lead to a global phase diagram
\cite{Si06.1,Si10.1,Col10.2,Voj08.2}, featuring both $J_{\rm K}$ (or $J_{\rm K}
N_{\rm F}$, see Box\,\ref{box_Kondo} in supplementary materials) and the
frustration parameter $G$ as non-thermal tuning parameters
(Fig.\,\ref{entropy}c). At large values of $G$, when the Kondo coupling is not
too large, the static Kondo singlet amplitude vanishes, and a paramagnetic phase
with small Fermi surface, P$_{\rm S}$, is stabilized, possibly in the form of a
metallic spin liquid. Concrete theoretical studies of the global phase diagram
in frustrated Kondo lattice models have been carried out, using a large-$N$
method in the case of the Shastry-Sutherland lattice \cite{Pix14.1} and through
a quantum Monte Carlo technique for a model on the honeycomb lattice
\cite{Sat18.1}.

To link such theoretical phase diagrams to experiments, one should be able to
quantify the degree of frustration and, ideally, continuously vary it by some
external tuning parameter. In insulating quantum magnets the frustration
strength can be quantified by the ratio of the paramagnetic Weiss temperature
and the ordering temperature or, in a more sophisticated way, by modelling
magnetic structure data to extract exchange interactions \cite{Kur19.1,Li20.1}.
In heavy fermion systems, however, the local moment exchange interaction is
typically dominated by the long-ranged RKKY interaction, and thus the conduction
electrons and anisotropies in their densities of states at the Fermi level may
also play an important role in defining the degree of frustration.

\begin{figure}[h!]
\centering
{\hspace{-0.2cm}\includegraphics[height=0.34\textwidth]{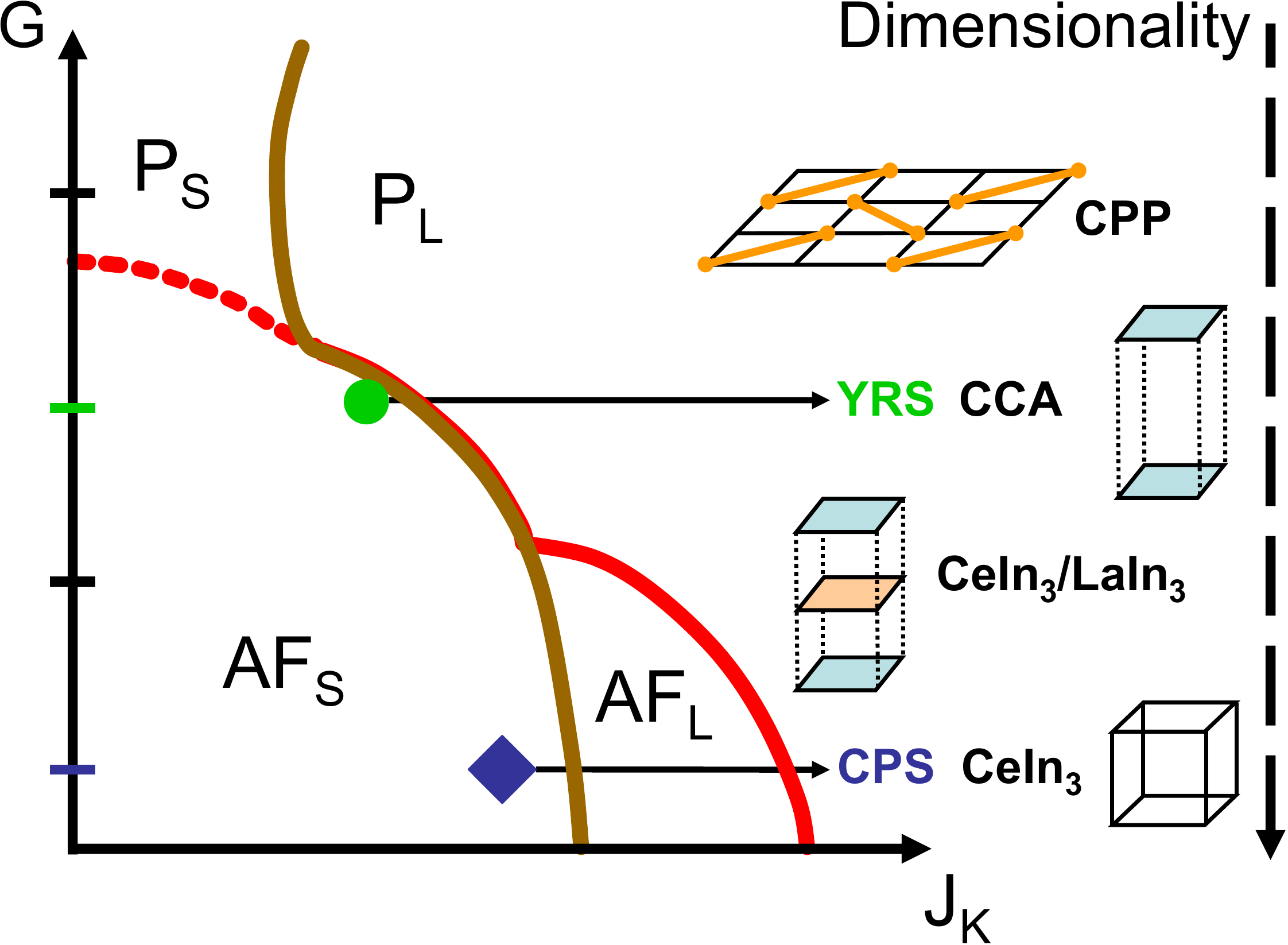}\hspace{1.5cm}\includegraphics[height=0.34\textwidth]{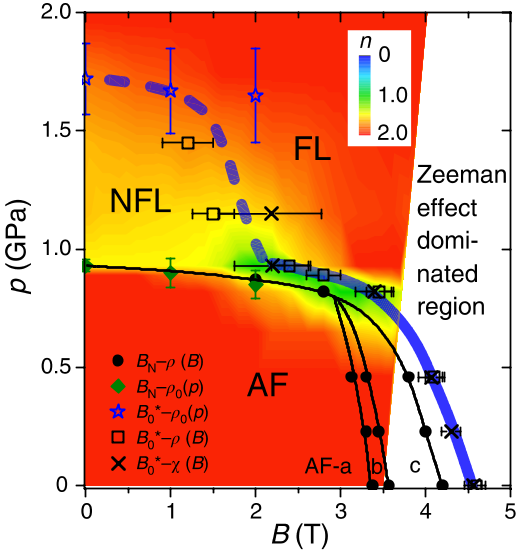}}
\vspace{0.5cm}

{\includegraphics[height=0.34\textwidth]{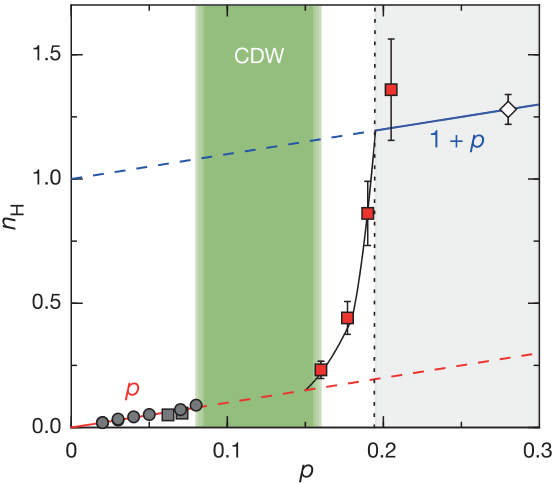}\hspace{1.5cm}\includegraphics[height=0.34\textwidth]{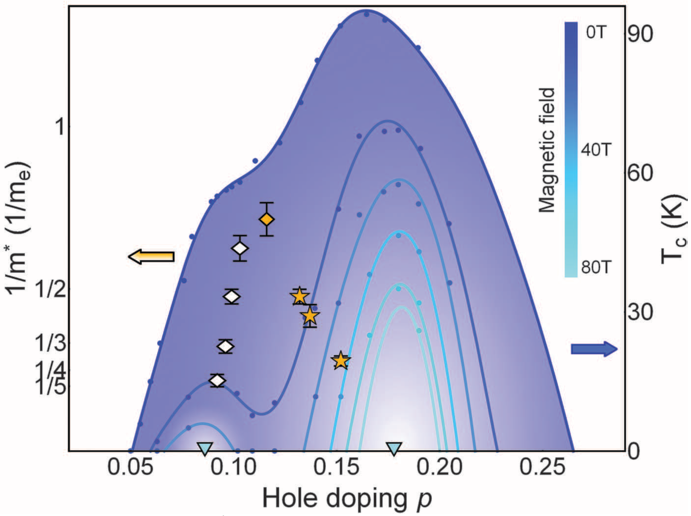}}
\vspace{0.5cm}

{\includegraphics[height=0.31\textwidth]{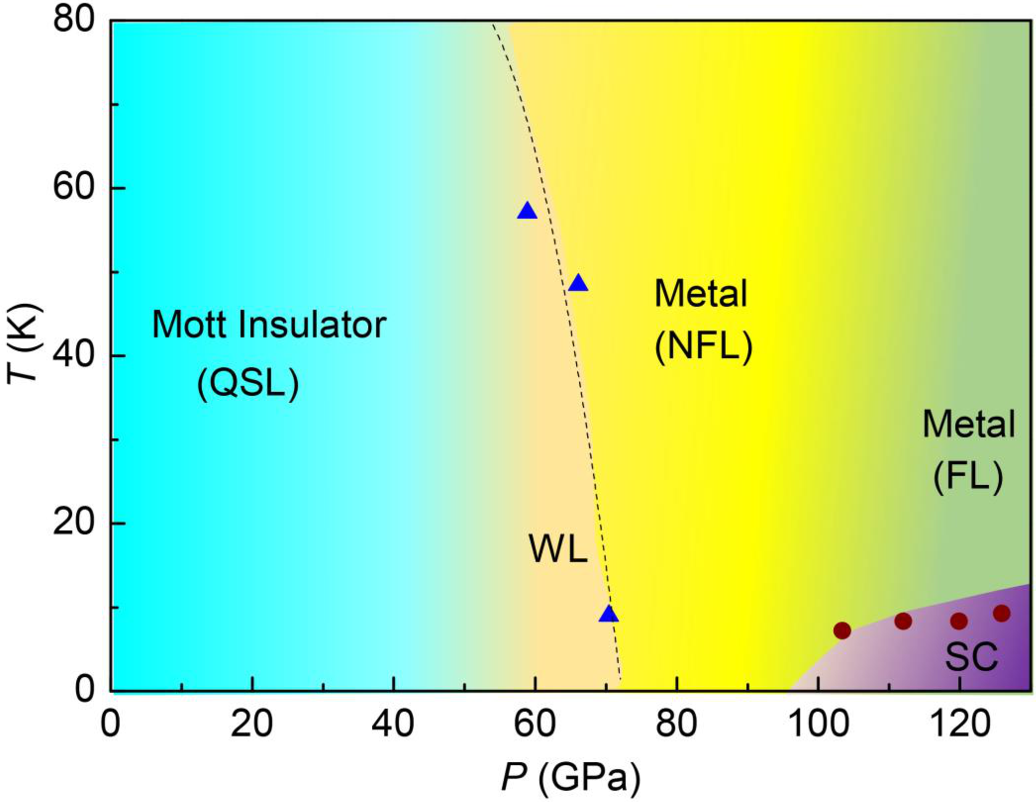}\hspace{0.3cm}\includegraphics[height=0.31\textwidth]{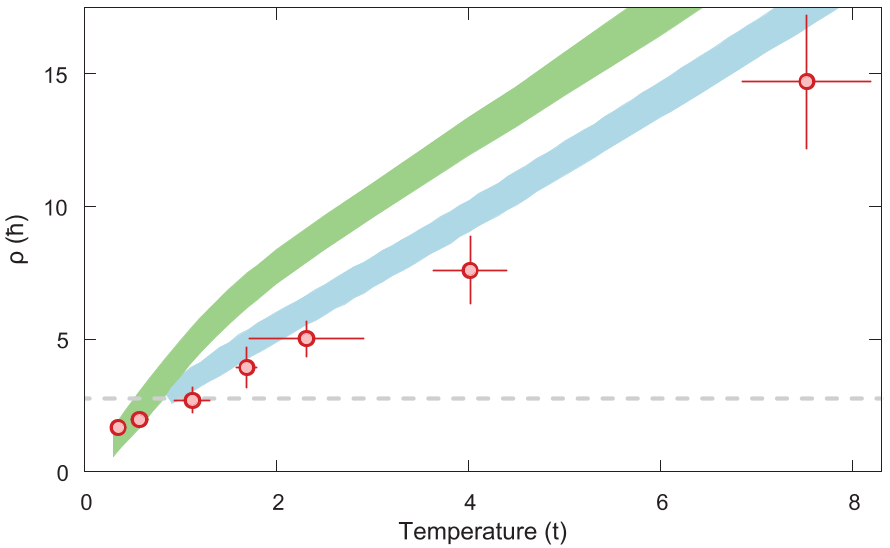}}
\vspace{0.5cm}

\vspace{-18.6cm}

\hspace{-6cm}{\bf\textsf{a}}\hspace{9cm}{\bf\textsf{b}}
\vspace{5.4cm}

\hspace{-7.7cm}{\bf\textsf{c}}\hspace{7.3cm}{\bf\textsf{d}}
\vspace{5.3cm}

\hspace{-8.1cm}{\bf\textsf{e}}\hspace{7cm}{\bf\textsf{f}}
\vspace{5.5cm}

\caption{\label{outreach} {\bf Broader implications.} {\bf a}$|$ Schematic
zero-temperature phase diagram with Kondo coupling $J_{\rm K}$ and frustration
parameter $G$ axes. The phases have the same meaning as in Fig.\,\ref{entropy}c.
Dimensionality (right) helps to calibrate the placement of selected materials
(CPS: Ce$_3$Pd$_{20}$Si$_6$, YRS: YbRh$_2$Si$_2$, CCA: CeCu$_{6-x}$Au$_x$, CPP:
Ce$_2$Pt$_2$Pb, reproduced with permission from ref.\citenum{Cus12.1}). {\bf
b}$|$ Pressure--magnetic field phase diagram of CePdAl, extrapolated from finite
temperature results to $T=0$. The colour code represents $n$ in $\Delta\rho \sim
A T^n$, revealing a range of non-Fermi liquid (NFL) behavior (adapted with
permission from ref.\citenum{Zha19.3}). {\bf c}$|$ Doping dependence of the Hall
number in the normal state for LSCO (circles),}
\end{figure}

Yet, so far, frustration in heavy fermion systems is discussed mostly in terms
of the more intuitive concept of local moments situated on (partially)
frustrated lattices
\cite{Nak06.1,Kim11.1,Fri14.1,Tok15.1,Wu16.2,Zha19.3,Kav20.1x}. Examples are
Pr$_2$Ir$_2$O$_7$ with a pyrochlore lattice of Pr atoms \cite{Nak06.1,Kav20.1x},
CeRhSn \cite{Tok15.1} and CePdAl \cite{Fri14.1,Zha19.3} with Ce atoms located on
distorted kagome planes, and HoInCu$_4$ with an fcc lattice of the Ho atoms
\cite{Sto20.1} (though the Kondo interaction in this low-carrier density system
is likely negligible). In Pr$_2$Ir$_2$O$_7$, non-Fermi liquid behavior is
observed above certain cutoffs in temperature and field (possibly due to spin
freezing) \cite{Nak06.1,Tok14.1}, and could be due to spin liquid behavior or
quantum criticality from a nearby magnetic QCP. In CeRhSn, thermodynamic
properties show non-Fermi liquid behavior only within the frustrated plane
\cite{Tok15.1}, and its suppression under uniaxial pressure applied within the
plane, which lifts the frustration \cite{Kue17.1}. This provides evidence for
frustration-induced non-Fermi liquid behavior. In CePdAl, a region of non-Fermi
liquid behavior was found in a pressure--magnetic field phase diagram, and
associated with the P$_{\rm S}$ phase \cite{Zha19.3} (Fig.\,\ref{outreach}b).
Regions of non-Fermi liquid behavior are, however, also seen in heavy fermion
metals without any obvious element of geometric frustration
\cite{Fri09.1,Cus10.1}. In YbRh$_2$(Si$_{1.95}$Ge$_{0.05}$)$_2$, a P$_{\rm S}$
phase was shown to be nested between an AF$_{\rm S}$ phase and a P$_{\rm L}$
phase \cite{Cus10.1}, thus directly confirming trajectory III in the theoretical
phase diagram (Fig.\,\ref{entropy}c). Further work is needed to understand how
this P$_{\rm S}$ phase relates to the ones hinted at in the above geometrically
frustrated local moment systems with much weaker Kondo interaction. A recent STM
study on Pr$_2$Ir$_2$O$_7$ provided evidence for a P$_{\rm L}$ to P$_{\rm S}$
transition through the tuning of minute disorder potential on the nanoscale
\cite{Kav20.1x}.

A new platform for the exploration of this physics are heavy fermion thin films.
In insulating local moment systems, quantum fluctuations are not only boosted by
(geometrical) frustration but also by reduced dimensionality
(Fig.\,\ref{entropy}d). Thus, in the search for materials in which the $G$
parameter can be (quasi-)continuously varied, also systems with tunable
dimensionality should be considered \cite{Cus12.1}. High-quality thin films
grown by molecular beam epitaxy \cite{Shi10.1,Ish16.1,Pro20.1} appear as
promising way forward. Superlattices of heavy fermion metals with normal metals
(CeIn$_3$/LaIn$_3$ and CeRhIn$_5$/YbRhIn$_5$) indeed showed the emergence of
non-Fermi liquid signatures with decreasing superlattice period
\cite{Shi10.1,Ish16.1}. Whether this represented the appearance of the P$_{\rm
S}$ phase (along trajectory III in Fig.\,\ref{entropy}c), or quantum criticality
from the suppression of antiferromagnetic order (along trajectory I in
Fig.\,\ref{entropy}c) is an interesting question for future studies, including
ideally also superlattices with passive (non-metallic) spacer compounds. In the
opposite direction, one can reduce $G$ by going to the three-dimensional limit.
The cubic compound Ce$_3$Pd$_{20}$Si$_6$ fits this prescription.
Magnetotransport and thermodynamic measurements have firmly established that the
Kondo destruction appears within the ordered part of the phase diagram
(Fig.\,\ref{hybrid}a-c and Fig.\,\ref{outreach}a) \cite{Cus12.1}.

\subsection{Entwined degrees of freedom and hybrid interactions}

We have so far considered the case of localized spins coupled to each other and
to conduction electrons by spin-exchange interactions. The rich settings of
condensed matter systems, however, allow us to utilize, or even design,
alternative degrees of freedom. In heavy fermion systems, physics beyond the
spin-only case comes in naturally: For the $4f$ ($5f$) electrons of the rare
earth (actinide) elements, the intraatomic spin-orbit coupling is large, and
thus having effective moments of total angular momentum $J>1/2$, that encompass
both spins (dipoles) and higher multipolar moments, is the generic case. Crystal
electric fields split these spin-orbit coupled states, but even the ground state
will, in general, not be a spin-only Kramers doublet.

\begin{figure}[h!]
\centering

\begin{minipage}[t]{0.5\textwidth}
\includegraphics[width=1\textwidth]{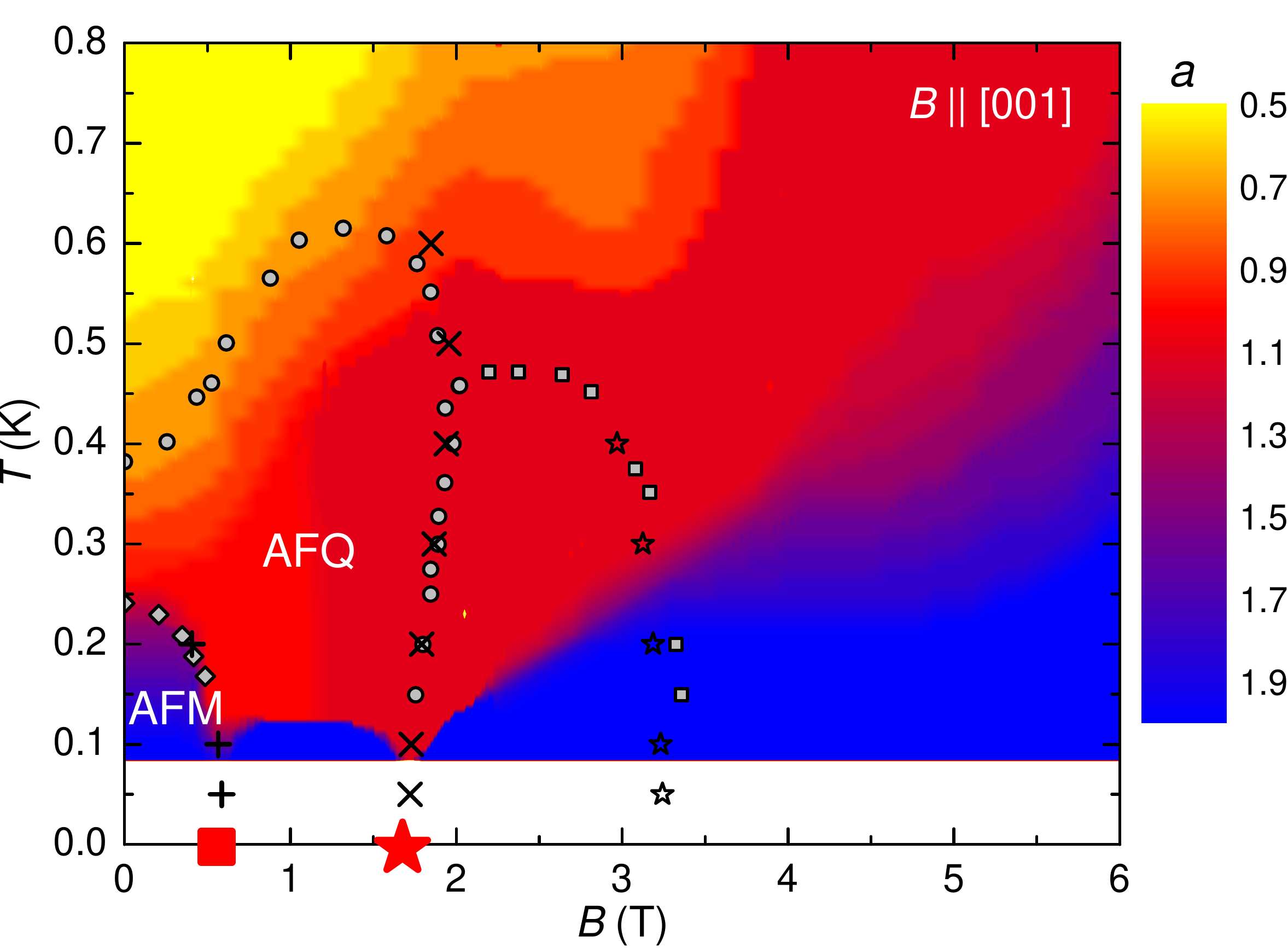}
\vspace{-0.3cm}

\hspace{0.3cm}\includegraphics[width=0.9\textwidth]{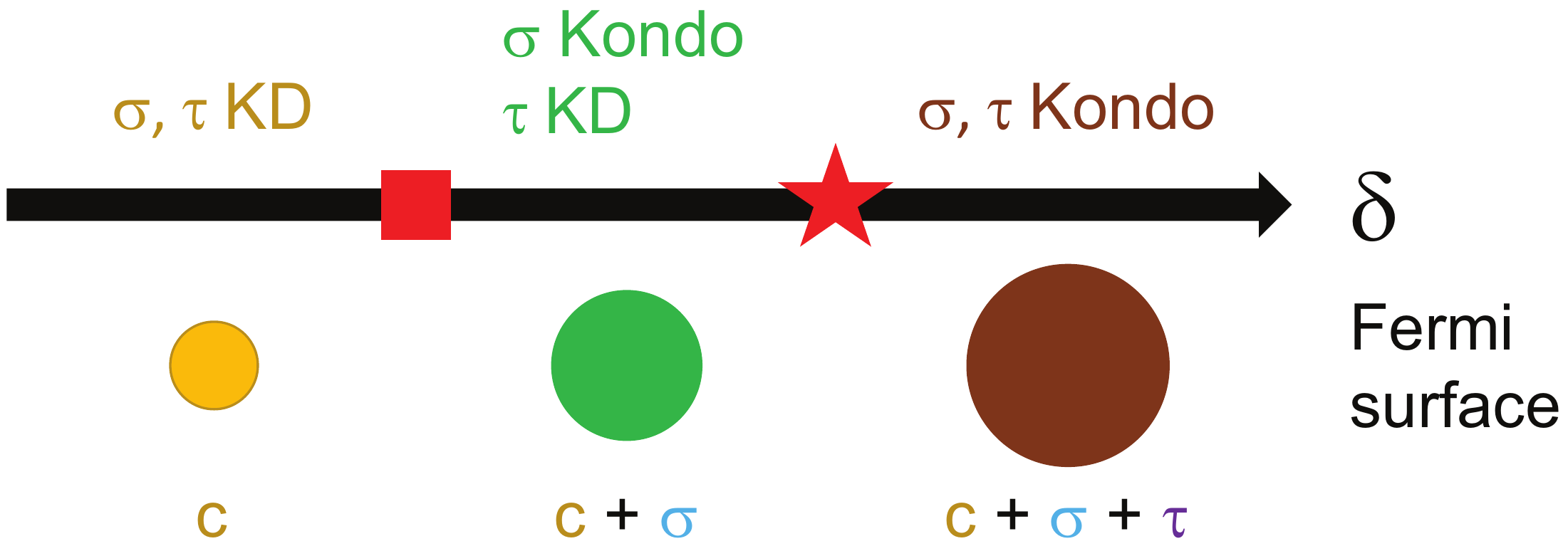}
\vspace{0.1cm}

\hspace{1cm}\includegraphics[width=0.44\textwidth]{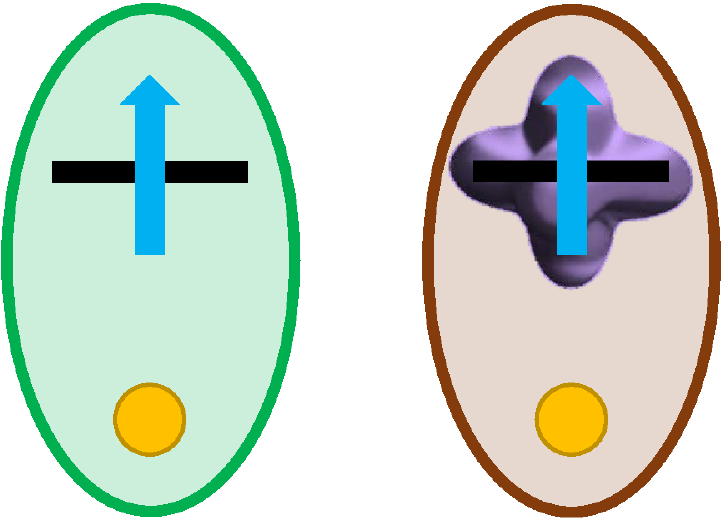}
\vspace{0.9cm}

\hspace{-0.4cm}\includegraphics[width=0.75\textwidth]{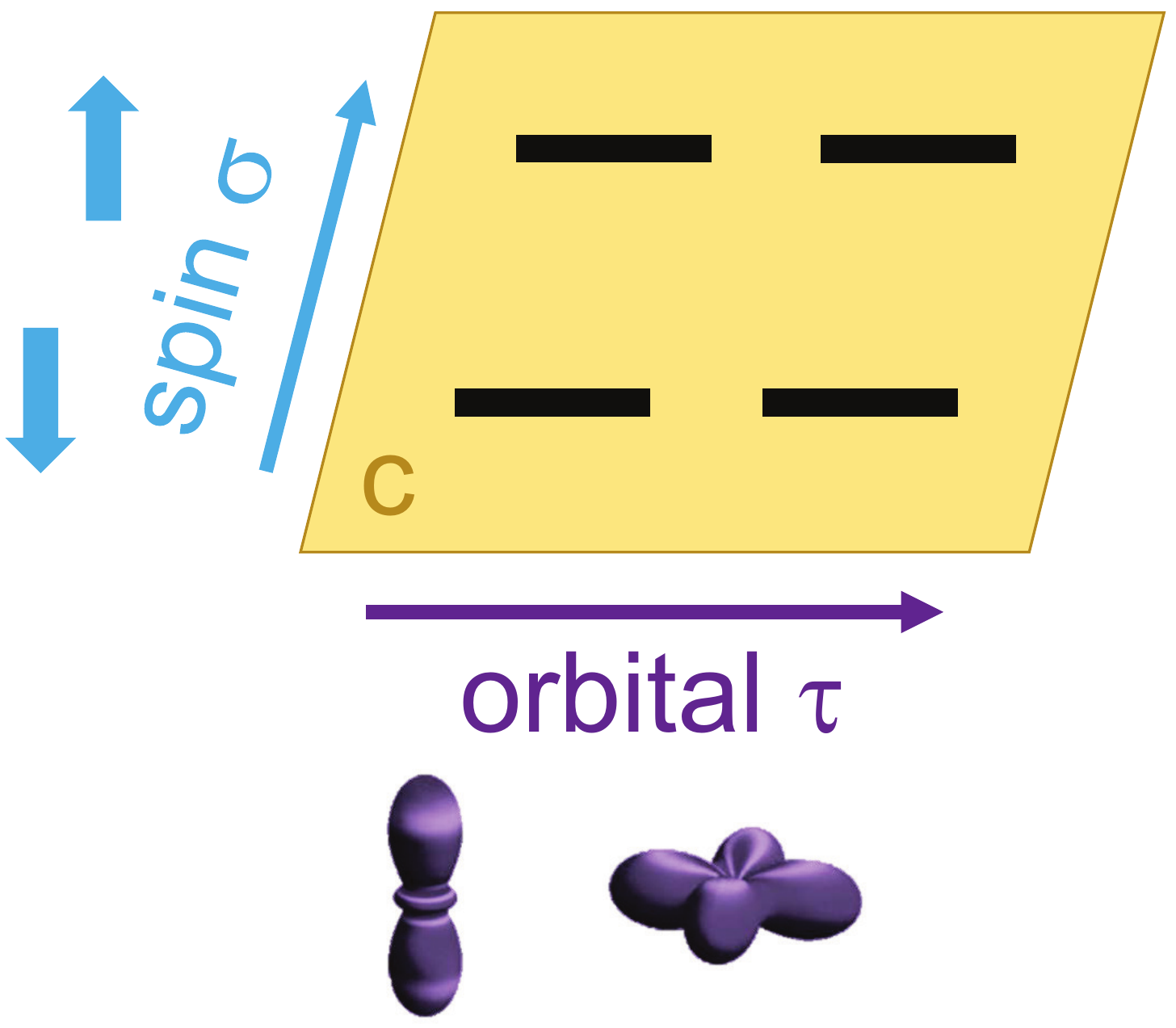}
\vspace{-18.7cm}

\hspace{-8.1cm}{\bf\textsf{a}}
\vspace{5.5cm}

\hspace{-8.1cm}{\bf\textsf{b}}
\vspace{5.5cm}

\hspace{-8.1cm}{\bf\textsf{c}}

\end{minipage}\hfill
\begin{minipage}[t]{0.5\textwidth}
\vspace{-5.8cm}

\includegraphics[width=0.62\textwidth]{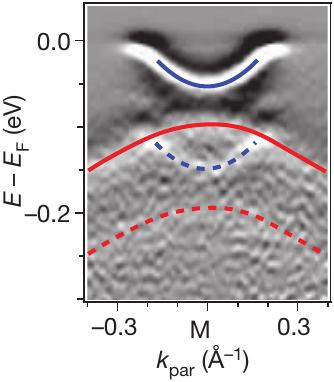}
\vspace{0.3cm}

\includegraphics[width=0.95\textwidth]{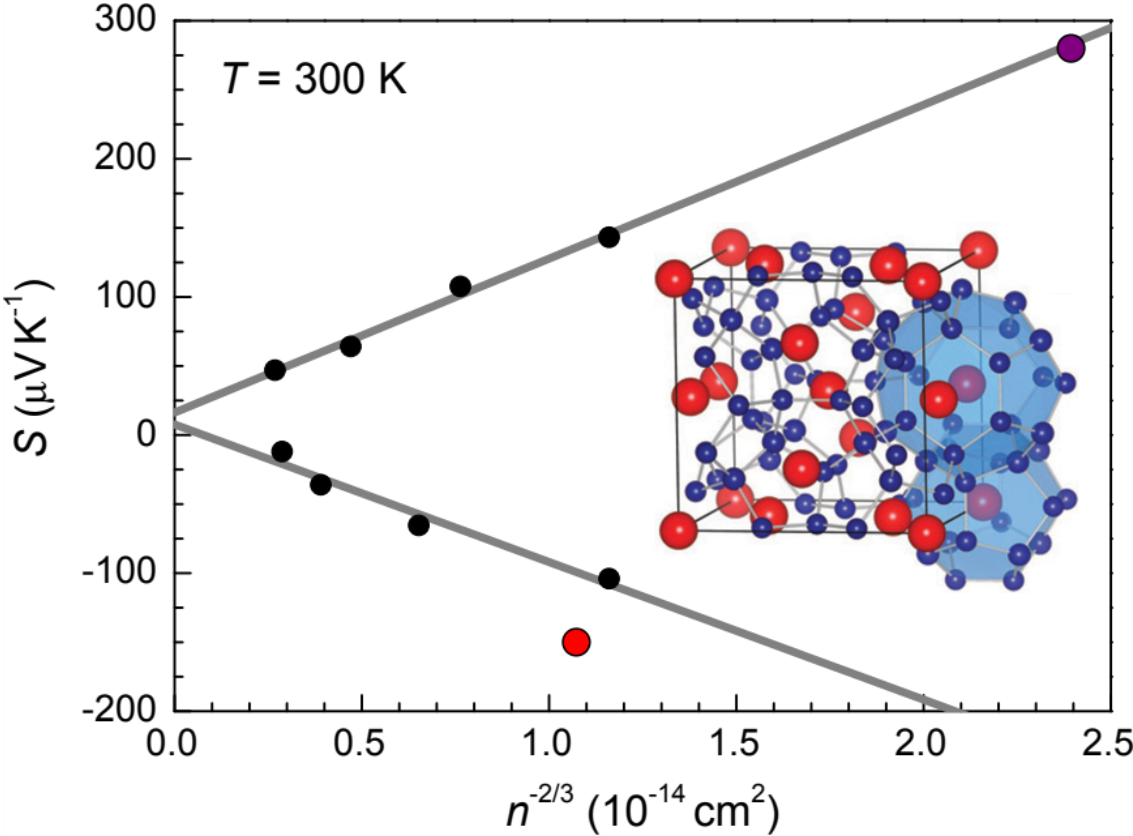}
\vspace{0.4cm}

\hspace{0.3cm}\includegraphics[width=0.68\textwidth]{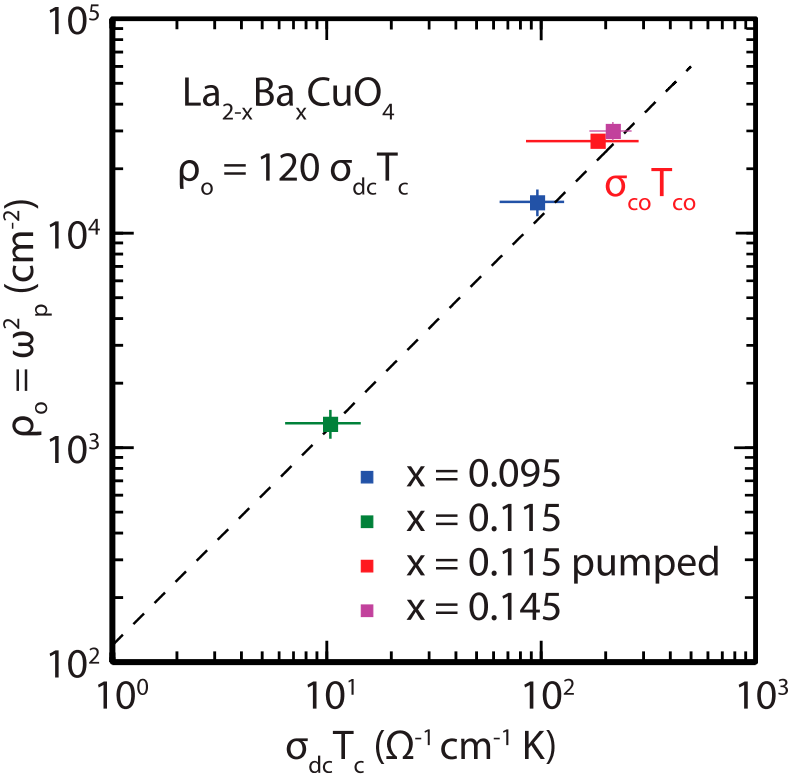}
\vspace{-18.6cm}

\hspace{-6cm}{\bf\textsf{d}}
\vspace{5.5cm}

\hspace{-7.3cm}{\bf\textsf{e}}
\vspace{5.5cm}

\hspace{-7.3cm}{\bf\textsf{f}}

\end{minipage}

\vspace{5.5cm}

\caption{\label{hybrid} {\bf Composite and boosted interactions.} {\bf a}$|$
Temperature-tuning parameter phase diagram of the heavy fermion compound
Ce$_3$Pd$_{20}$Si$_6$ (adapted with permission from ref.\citenum{Mar19.1}). {\bf
b}$|$ Sketches visualizing the interaction of spin ($\sigma$) and orbital
($\tau$) degrees of freedom with conduction electrons (c), in the form of two
stages of Kondo destruction (KD, adapted with permission from
ref.\citenum{Mar19.1}). {\bf c}$|$ Degrees of freedom that entangle in the Kondo
phases of Ce$_3$Pd$_{20}$Si$_6$, adapted with permission from
ref.\citenum{Mar19.1}. {\bf d}$|$ ARPES image of monolayer FeSe on SrTiO$_3$,
revealing replica (dashed guides-to-the-eyes) of the main bands}
\end{figure}

An example is the cubic heavy fermion compound Ce$_3$Pd$_{20}$Si$_6$. Its
Ce$4f^1$ crystal-field-split ground state is a $\Gamma_8$ quartet, which has
multipolar character. Surprisingly simple low-energy behavior was observed
\cite{Mar19.1}: Across two QCPs, one at the border of antiferromagnetic (AFM),
the other at the border of antiferroquadrupolar (AFQ) order
(Fig.\,\ref{hybrid}a), two distinct electron localization transitions take
place, driven by a single degree of freedom at a time (Fig.\,\ref{hybrid}b,c).
They were understood as a sequential destruction of an SU(4)
spin-orbital-coupled Kondo effect \cite{Mar19.1}.

Also in other classes of strongly correlated electron systems there is ample
evidence for entwined degrees of freedom. In the manganites \cite{Tok00.1} and
fullerides \cite{Tak09.1}, it is also spin and orbital degrees of freedom that
interplay, and in the iron pnictides and chalcogenides, more than one $3d$
orbital is important for the physics near the Fermi energy \cite{Si16.1}. In the
cuprates, charge order emerges and interplays with the spin degrees of freedom
\cite{Bad16.1,Ram15.1}, and even in magic-angle bilayer graphene, the physics
likely depends on both the spin and valley degrees of freedom \cite{Cao18.1}. To
advance the understanding of the rich physical phenomena present in all these
materials, it is important to decipher the roles the different degrees of
freedom play in determining the stable phases (Fig.\,S1 in supplementary
materials) and excitations associated with them, as well as with (quantum) phase
transitions between them.

Strong electron interactions can also be modified by other interactions. In
fact, the competition between the different energy scales associated with the
various degrees of freedom---that also leads to the rich phase diagrams
(Fig.\,S1 in supplementary materials)---makes these systems generally highly
responsive to all kinds of stimuli. Particularly interesting are situations
where alternative interactions boost strong-correlation effects, thereby
enhancing functionality.

One example is ``phonon boosting''. The interplay of the electron-electron
Coulomb interaction and phonon-mediated electron interaction effects is captured
by the Anderson-Holstein model \cite{Hew02.1}, which has been much explored in
the context of single molecular transistors \cite{Lue08.1,Kal19.1}. In Kondo
systems, increasing the coupling of the conduction electrons with local optical
phonons was shown to enhance the Kondo temperature by orders of magnitude, up to
a point beyond which the charge Kondo effect is stabilized \cite{Hot07.1}. This
(former) mechanism was suggested to explain a high-temperature Kondo effect in a
Ce-based thermoelectric ``rattler'' compound \cite{Pro13.1}. Its thermopower at
room temperature was shown to be strongly enhanced over values found in a $4f$
moment-free reference compound (Fig.\,\ref{hybrid}e) though the bare Kondo
temperature of the system at low temperatures, where the rattling mode is not
activated, was only 1\,K. This effect almost doubled the thermoelectric figure
of merit \cite{Pro13.1}. The strong response to phonons in this type-I clathrate
compound is due to its special crystal structure, with an atom trapped in an
oversized cage. Whereas most heavy fermion compounds lack such special structure
elements and are much less sensitive to phonons, other strongly correlated
electron systems, in particular with polar bonds and/or reduced dimensionality,
do show phonon- (as well as photon-)boosting effects; these will be discussed in
Sect.\,\ref{implications}.

\subsection{Correlation-driven topology}

With the advent of ``topology'' in electronic materials
\cite{Hal17.1,Has10.1,Qi11.1,NatMater16.1,Ban16.2,Wen17.1,Arm18.1} also the
question of how strong correlations and topology may interplay is attracting
much interest. In light of recent developments, we here focus on Weyl semimetals
\cite{NatMater16.1,Ban16.2,Arm18.1} (Box\,\ref{box_topology} in supplementary
materials).

Theoretical predictions for noninteracting electron systems are becoming ever
more efficient in guiding the search for new candidate materials. In principle,
topological invariants---key quantities in defining the topological nature of 
the electrons' wavefunctions---can be directly calculated from electronic band
structure. Because this is a laborious process, other, higher throughput methods
were also developed. These classify materials by certain indicators for
topology, most notably symmetry eigenvalues
\cite{Chi16.1,Po17.1,Zha19.1,Ver19.1,Tan19.1}.

Despite---or even because---these approaches have by now identified enormous
numbers of candidate materials, clear-cut experimental confirmations remain a
formidable task. ARPES is playing a central role in it (see panels a-d in
Box\,\ref{box_topology} in supplementary materials), even though the
interpretation is not always easy. On one hand, to disentangle surface from bulk
and topological from topologically trivial states, comparison with {\em ab
initio} calculations is needed, which are challenging for real surfaces even in
the noninteracting case. On the other hand, the rather intense incident light
may ``dope'' the surface and thus modify the bandstructure \cite{Fra17.1}.
Complementary experimental techniques, including (magneto)transport, optical
spectroscopy, and quantum oscillation experiments, are thus equally important,
and have indeed provided independent evidence
\cite{Hua15.2,Mol16.1,Zha16.3,Arm18.1,Zha19.2}.

\begin{figure}[t!]
\centering
\begin{minipage}[t]{0.48\textwidth}
\includegraphics[width=1.01\textwidth]{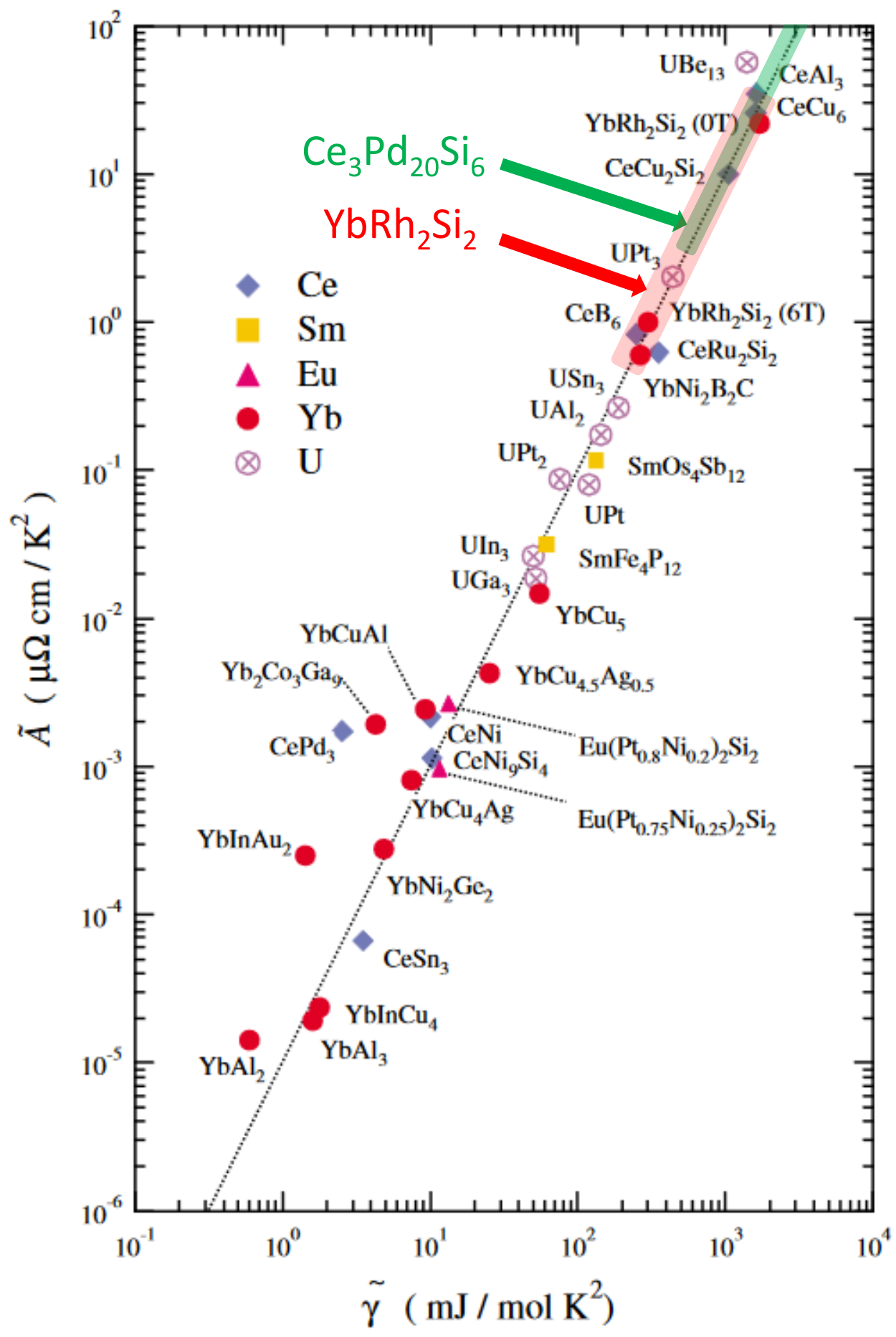}
\end{minipage}
\begin{minipage}[t]{0.45\textwidth}
\vspace{-11.82cm}

\hspace{0.01cm}\includegraphics[width=1.03\textwidth]{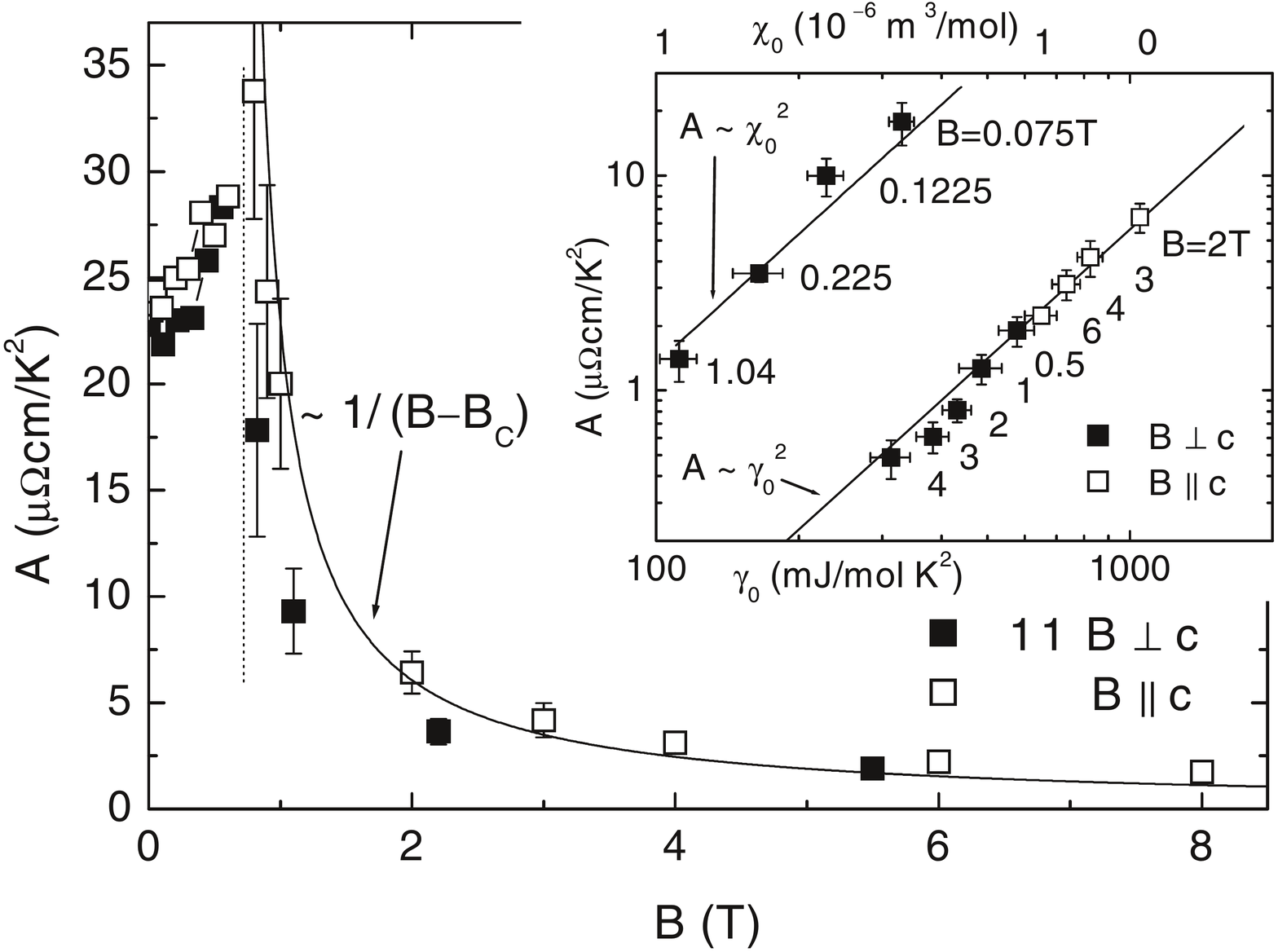}\\[0.04cm]
\hspace{0.12cm}\includegraphics[width=0.98\textwidth]{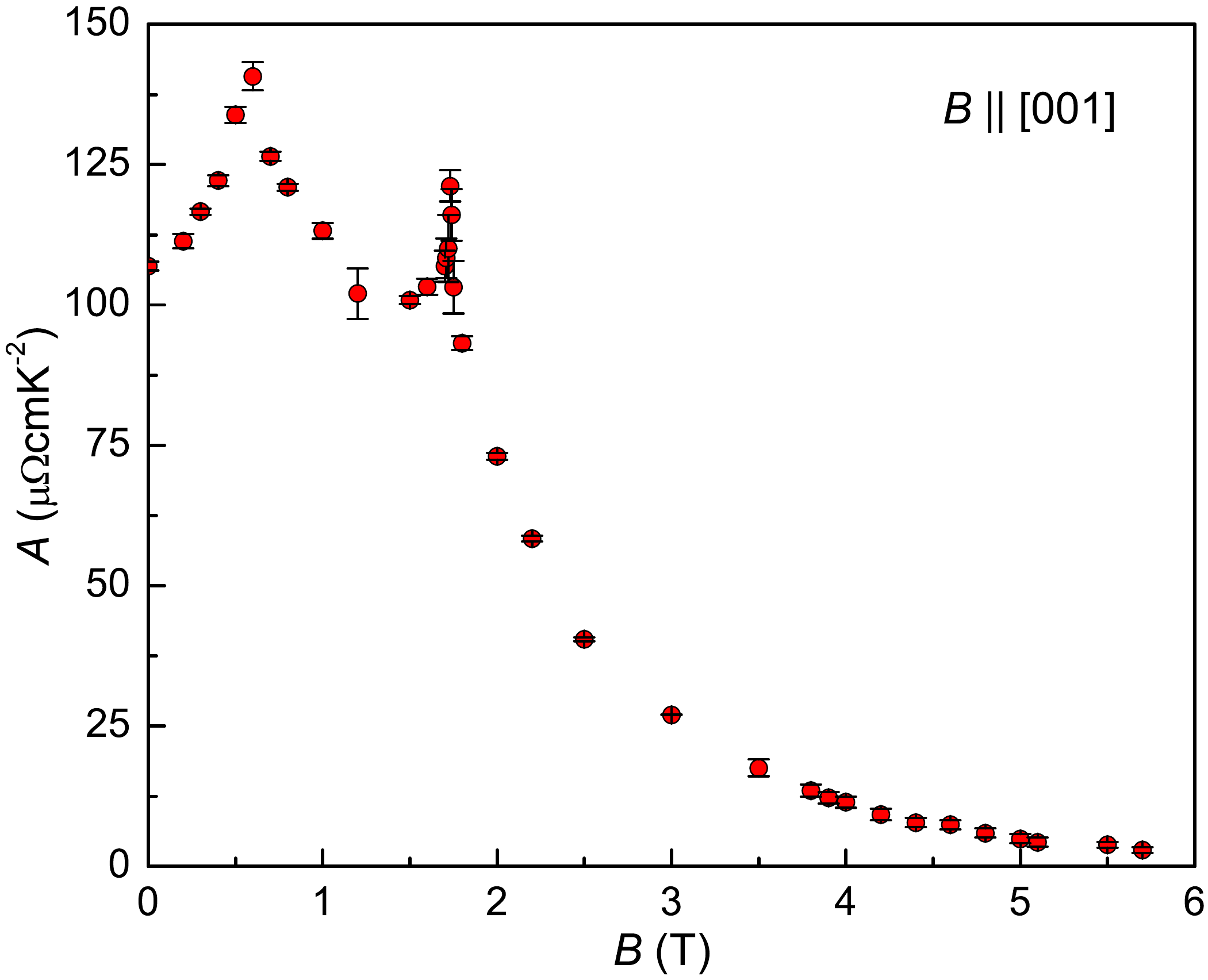}\hfill

\vspace{-11.34cm}

\hspace{-1.8cm}\textcolor{red}{\textsf{YbRh$_2$Si$_2$}}
\vspace{5.4cm}

\hspace{4.7cm}\textcolor{darkgreen}{\textsf{Ce$_3$Pd$_{20}$Si$_6$}}
\vspace{4.5cm}

\end{minipage}
\vspace{-12.3cm}

\hspace{-6.9cm}{\bf\textsf{a}}\hspace{7.8cm}{\bf\textsf{b}}
\vspace{4.7cm}

\hspace{1.2cm}{\bf\textsf{c}}
\vspace{6.2cm}

\begin{minipage}[t]{0.43\textwidth}
\includegraphics[width=0.95\textwidth]{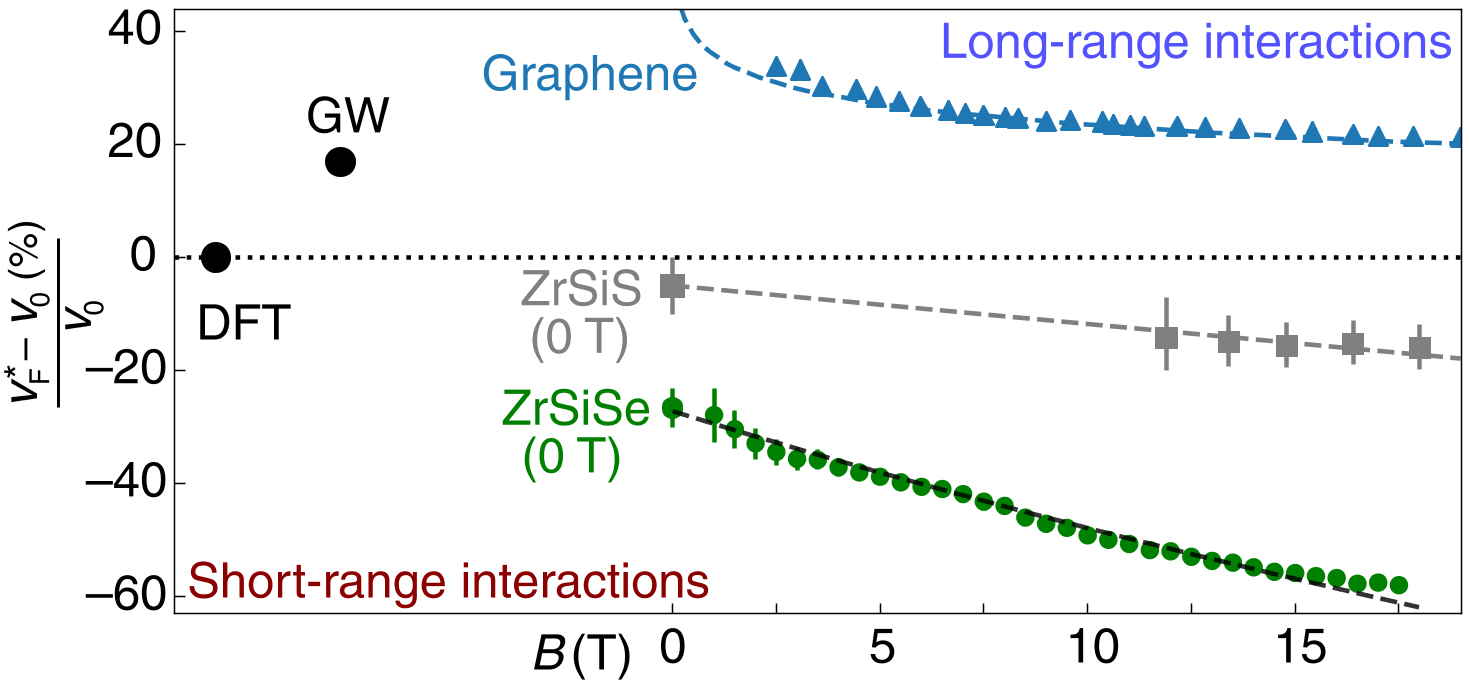}
\end{minipage}
\begin{minipage}[t]{0.5\textwidth}
\vspace{-3.14cm}

\hspace{0.3cm}\includegraphics[width=0.95\textwidth]{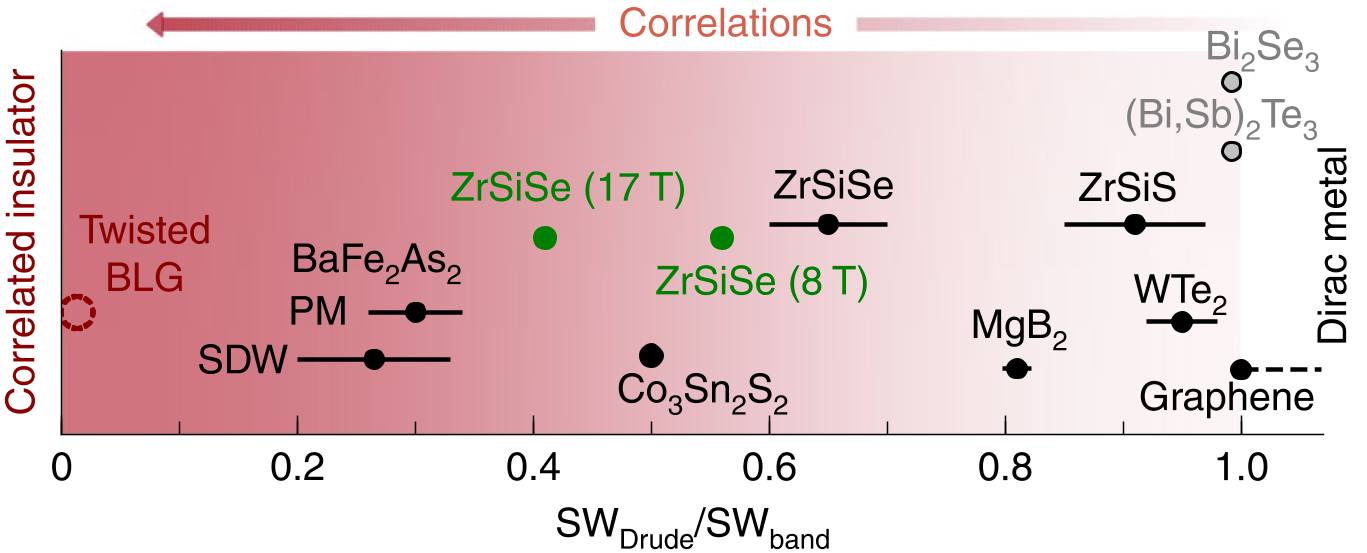}
\end{minipage}
\vspace{-3.7cm}

\hspace{-7.7cm}{\bf\textsf{d}}\hspace{7cm}{\bf\textsf{e}}
\vspace{3.5cm}

\caption{\label{tuning} {\bf Tuning correlation strength.} {\bf a}$|$
Generalized Kadowaki-Woods plot, adapted with permission from
ref.\citenum{Tsu05.1} to include ranges of $A$ coefficient reached in
experiments on the heavy fermion compounds YbRh$_2$Si$_2$ from panel {\bf b} and
Ce$_3$Pd$_{20}$Si$_6$ from panel {\bf c} as shadings. {\bf b}$|$ $A$ coefficient
of the low-temperature electrical resistivity of YbRh$_2$Si$_2$ as function of
applied magnetic field $B=\mu_0 H$. The inset relates the $A$ coefficient to the
Sommerfeld coefficient $\gamma_0$ and the Pauli susceptibility $\chi_0$ and
confirms the validity of the Kadowaki-Woods and Sommerfeld-Wilson ratios
(adapted with permission from ref.\citenum{Geg02.1}). {\bf c}$|$ $A$ coefficient
of the low-temperature electrical resistivity of Ce$_3$Pd$_{20}$Si$_6$ vs $B$,
evidencing mass enhancements at two distinct fields (adapted with permission
from ref.\citenum{Mar19.1}). {\bf d} Fermi velocity renormalization with respect
to the noninteracting value from DFT vs magnetic field for the nodal-line
semimetal ZrSiSe, the nodal-loop semimetal ZrSiS,}
\end{figure}

To address the effect of correlations, theoretical treatments can either (i)
study how electron correlations modify (known) noninteracting topological
states, or (ii) explore how nontrivial topology can be introduced into (known)
strongly correlated electronic states. A recent example in the spirit of
approach (i) is an optical spectroscopy study of the nodal-line semimetal ZrSiSe
\cite{Sha20.1}, where a comparison with density functional theory (DFT) revealed
a reduction of the Fermi velocity with respect to the noninteracting one derived
by DFT by almost 30\%, an effect that was attributed to interactions
(Fig.\,\ref{tuning}d). Similar effects are seen also in several other
topological semimetals (Fig.\,\ref{tuning}e); and also for graphene, a quantum
Monte Carlo study showed short- and long-range Coulomb interaction effects to
reduce the Fermi velocity by about 40\% near the strong-coupling fixed point
\cite{Tan18.1}.

As for Schr\"odinger particles, where even Fermi liquids with extreme
renormalizations (by orders of magnitude, see Fig.\,\ref{tuning}a) may---but do
not have to be (Fig.\,\ref{tuning}b,c)---adiabatically connected to the
noninteracting state, one could conceive such a situation also for Dirac
particles. Thus, approach (ii) may well discover phenomena that have no
analogues in the noninteracting world, a particularly exciting prospect.

\begin{figure}[h!]
\centering

{\includegraphics[height=0.34\textwidth]{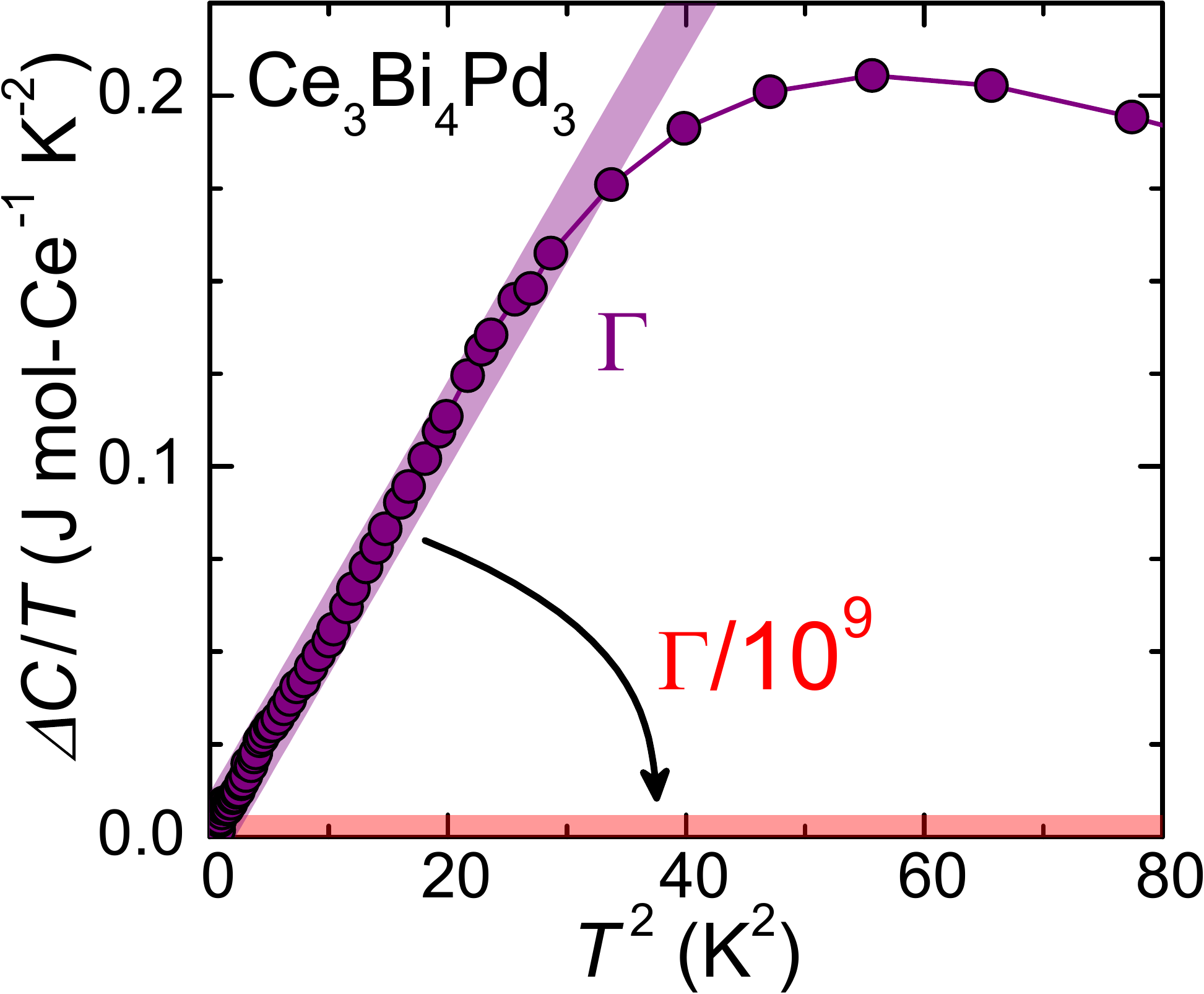}\hspace{1cm}\includegraphics[height=0.34\textwidth]{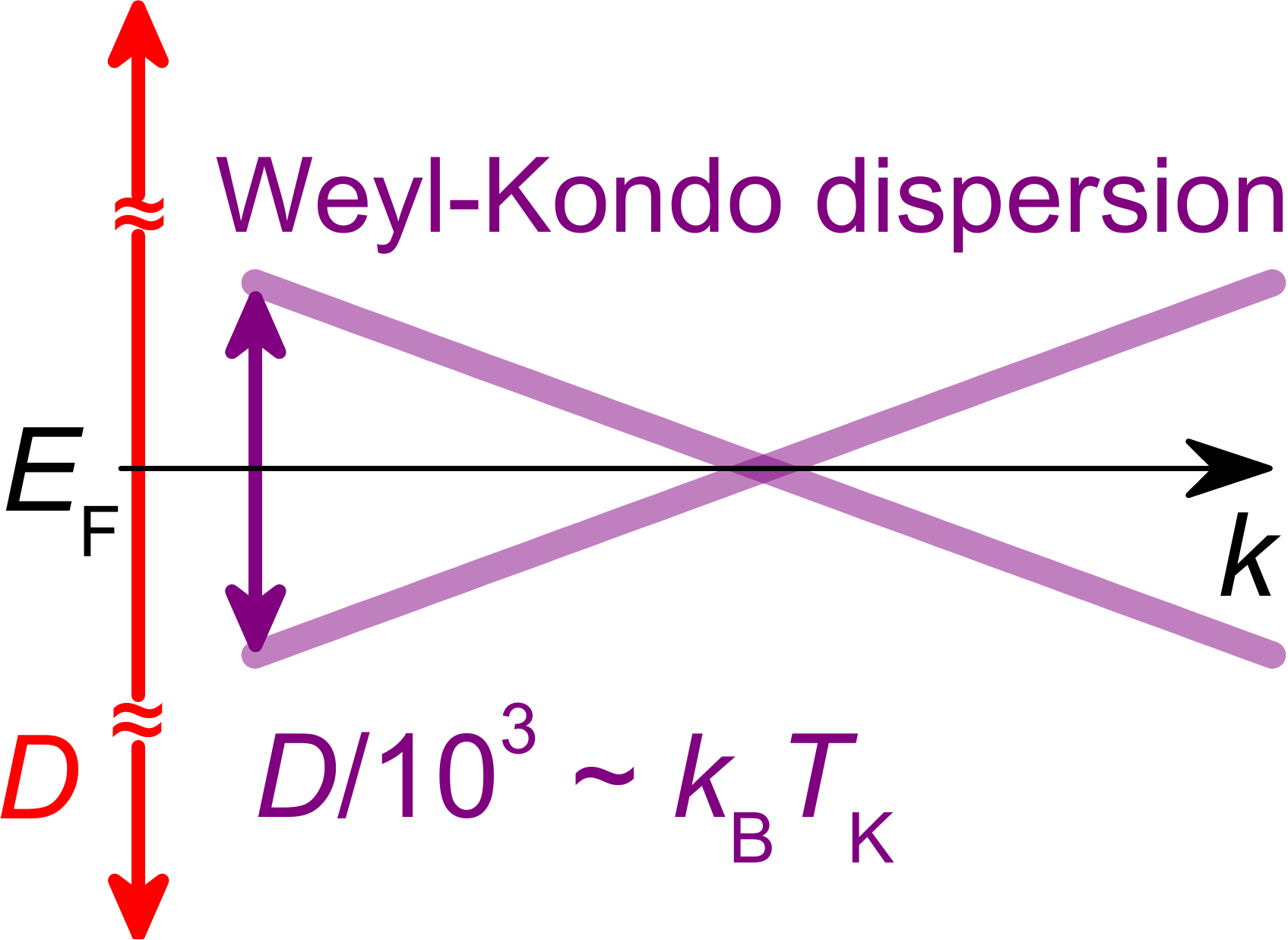}}
\vspace{0.6cm}

{\includegraphics[height=0.34\textwidth]{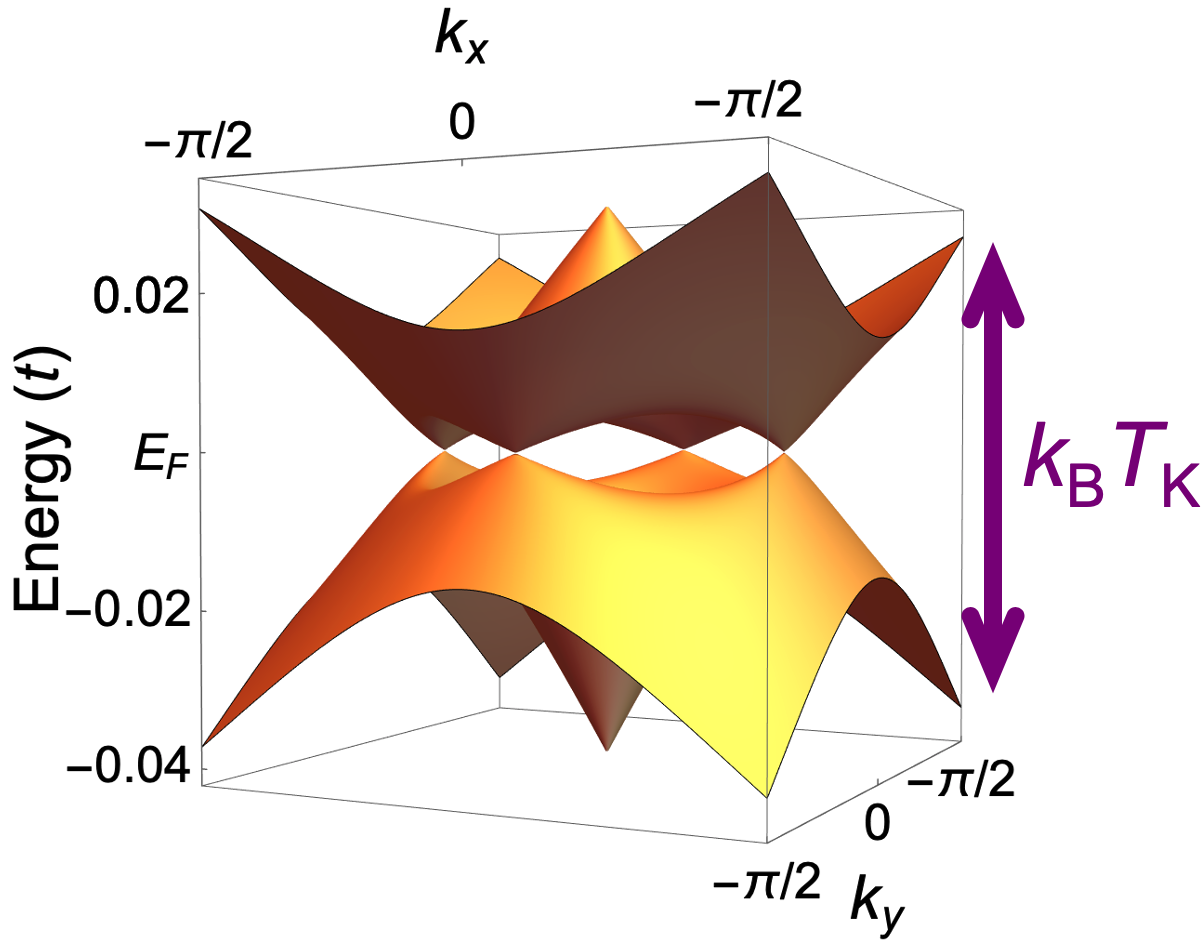}\hspace{1cm}\includegraphics[height=0.34\textwidth]{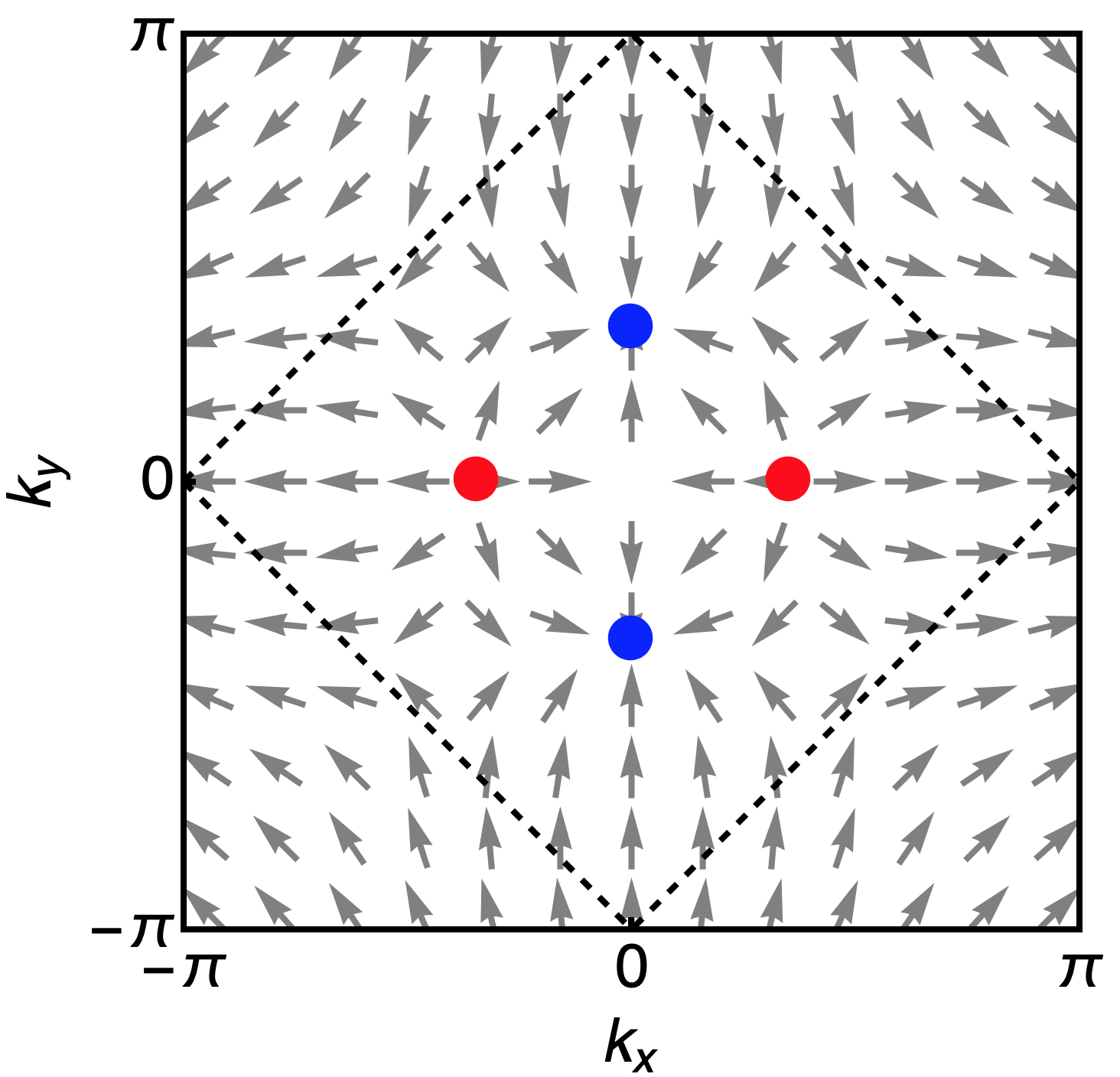}}
\vspace{0.5cm}
\vspace{-13.1cm}

\hspace{-7.9cm}{\bf\textsf{a}}\hspace{7.5cm}{\bf\textsf{b}}
\vspace{5.6cm}

\hspace{-5.9cm}{\bf\textsf{c}}\hspace{7.9cm}{\bf\textsf{d}}
\vspace{6.2cm}

\caption{\label{WeylKondo} {\bf Weyl-Kondo semimetal.} {\bf a}$|$ Electronic
specific heat coefficient $\Delta C/T$ of Ce$_3$Bi$_4$Pd$_3$, displaying
linear-in-$T^2$ behavior at low temperatures (adapted with permission from
ref.\citenum{Dzs17.1}). {\bf b}$|$ Sketch of the linear electronic dispersion in
momentum space near a Weyl point and the band renormalization corresponding to
the specific heat result. {\bf c}$|$ Dispersion of a theoretical model for a
Weyl-Kondo semimetal (adapted with permission from ref.\citenum{Lai18.1}). {\bf
d}$|$ Berry curvature field near the Weyl and anti-Weyl nodes (red and blue
dots, respectively) for the same model (adapted with permission from
ref.\citenum{Lai18.1}).}
\end{figure}

A promising setting to find such entirely new electronic phases are systems
where strong electron correlations and large spin--orbit interaction coexist, as
is the case in heavy fermion compounds. To single out the effect of the
spin--orbit interaction, a strategy to selectively tune its strength should be
identified. This attempt was made in a chemical substitution study of the Kondo
insulator Ce$_3$Bi$_4$Pt$_3$ where the $5d$ element Pt was successively replaced
by the $4d$ element Pd \cite{Dzs17.1}. As this substitution is isostructural,
isoelectronic, and essentially isosize, the large mass difference between Pt and
Pd and the associated difference in atomic spin--orbit coupling strength was
suggested to play the dominant role in the tuning \cite{Dzs17.1}. A
transformation from an insulator to a semimetal was observed with increasing Pd
content, with the end compound Ce$_3$Bi$_4$Pd$_3$ showing striking signatures of
nontrivial topology: The low-temperature electronic specific heat coefficient
$\Delta C/T$ is linear in $T^2$ (Fig.\,\ref{WeylKondo}a), pointing to a linear
electronic dispersion in momentum space, as expected for Dirac or Weyl fermions.
The velocity of the corresponding fermions calculated from the slope $\Gamma$ is
less than 1000\,m/s, which is three orders of magnitude lower than a typical
Fermi velocity of simple metals. Because this unusual specific heat contribution
appeared only below the Kondo temperature---which is again one thousandth of a
typical Fermi temperature---it was argued to be Kondo-driven \cite{Dzs17.1}.

Indeed, a theoretical study of the periodic Anderson model on a
noncentrosymmetric lattice discovered a Kondo-driven topological semimetal
phase, dubbed Weyl-Kondo semimetal \cite{Lai18.1}, with extremely flat linear
dispersion around Weyl nodes pinned to the Fermi energy
(Fig.\,\ref{WeylKondo}b,c), in agreement with the above experiments
\cite{Dzs17.1}. The Kondo effect cooperates with the nonsymmorphic space group 
of the underlying lattice, not only to generate the strongly renormalized Weyl
nodes but also to place them at the Fermi energy \cite{Gre20.1}. Weyl and
anti-Weyl nodes are sources and sinks of Berry curvature
(Fig.\,\ref{WeylKondo}d). Recent experiments have directly evidenced them via
spontaneous (zero magnetic field) Hall effect measurements \cite{Dzs18.1x}, and
even seen them annihilate in large magnetic fields \cite{Dzs19.1x}. The giant
magnitude of the spontaneous Hall effect, as well as its electric field and
frequency dependence were understood as the Weyl nodes being part of the Kondo
resonance, situated in immediate proximity to the Fermi energy \cite{Dzs18.1x}.
The calculations \cite{Lai18.1} also predicted surface states that feature 
strongly renormalized Fermi arcs, which still await experimental confirmation.

Weyl physics is being explored also in other Ce- and Yb-based intermetallic
compounds \cite{Guo17.2,Guo18.1,Sch18.1}, creating the exciting opportunity to
discover signatures of strong correlation-driven electronic topology also there.
Together with theoretical efforts
\cite{Zha16.4,Xu17.1,Roy17.1,Cha18.2,Iva19.1,Lu19.1,Wan19.1,Yan20.1,Gre20.1,Gre20.2}
this may help to establish the new field of strongly correlated electronic
topology.

%%%%%%%%%%%%%%%%%%%%%%%%%%%%%%%%%%%%%%%%%%%%%%%%%%%%%%%%%%%%%%%%%%%%%%%%%%%%%%
%%%%%%%%%%%% SECTION VI %%%%%%%%%%%%%%%%%%%%%%%%%%%%%%%%%%%%%%%%%%%%%%%%%%%%%%
%%%%%%%%%%%%%%%%%%%%%%%%%%%%%%%%%%%%%%%%%%%%%%%%%%%%%%%%%%%%%%%%%%%%%%%%%%%%%%

\section{Broader implications}\label{implications}

We have highlighted how the heavy fermion systems provide a setting to search
for and explore quantum criticality and novel quantum phases. Ultimately, the
field of strongly correlated systems aims to develop the organizing principles
that may operate across the materials platforms. With that consideration in
mind, we discuss how the insights gained from the heavy fermion field impact on
other classes of strongly correlated systems and beyond. 

\subsection{Quantum criticality and electronic localization--delocalization}\label{implicationsA}

In heavy fermion quantum criticality, the physics of Kondo destruction plays an
essential role. From the critical phenomenon perspective, the Kondo destruction
QCP involves critical physics that is beyond the Landau framework of order
parameter fluctuations. The destruction of the static Kondo effect gives rise to
critical modes that are in addition to the slow fluctuations of the
antiferromagnetic order parameter. From the correlated electron perspective,
this type of QCP epitomizes metallic quantum criticality in which the
suppression of electronic orders entwines with an electronic
localization--delocalization transition
\cite{Sch00.1,Pas04.1,Shi05.1,Fri10.2,Cus12.1,Pro20.1}. Such a transition of
electrons in a metallic background, and its accompanying jump of Fermi surface,
have also been emphasized in the context of hole-doped high $T_c$ cuprates
\cite{Ram15.1,Bad16.1}.

This analogy extends to the domain of theoretical analyses. In heavy fermion
systems, the destruction of the static Kondo effect and the associated jump of
the Fermi surface have been extensively analyzed using the EDMFT approach to the
Kondo lattice model \cite{Si01.1,Gre03.2,Hu20.1x}. A similar method has recently
been adopted to describe the localization--delocalization transition in the
single-band Hubbard model in the context of the high $T_c$ cuprates
\cite{Jos20.1,Cha20.1}. 

The global phase diagram of heavy fermion systems delineates whether a jump of
Fermi surface is concurrent with---or detached from---the suppression of
antiferromagnetic or other electronic orders. Also in the case of high $T_c$
cuprates, the relationship between a sudden reconstruction of the Fermi surface
and the development of electronic orders at zero temperature is a central
issue. 

Additionally, the localization--delocalization transition comes into play near
the Mott transition in Mott-Hubbard systems, as has been considered
theoretically \cite{Sen08.1,Ter11.1}. Organic charge-transfer salts have
provided a versatile materials setting to study this issue. For example,
measurements of the Hall coefficient in a doped organic superconductor,
$\kappa$-ET$_4$Hg$_{2.89}$Br$_8$, have hinted at a drastic Fermi surface 
transformation \cite{Oik15.1}. In intermetallic systems, there has been
considerable recent development on materials with a potential spin-liquid ground
state. Pressurizing such compounds presents a promising new setting to realize a
Mott transition and elucidate the accompanying evolution of the Fermi surface.
Progress along this direction has recently been reported in a high-pressure
study of NaYbSe$_2$ (Fig.\,\ref{outreach}e) \cite{Jia20.1}.

The Kondo destruction QCP, with the localization--delocalization transition of
the $f$ electrons in a metallic background \cite{Si01.1,Col01.1,Pep08.1},
represents the earliest example of orbital-selective Mott transitions. In $d$
electron systems, this transition was first discussed in the context of the
ruthenates \cite{Ani02.1}. Unlike in the heavy fermion case, the analysis was
done in the band basis; different bands, by definition, do not hybridize with
each other. In reality, the Coulomb interactions are expressed in the orbital
basis, and crystalline symmetry generically allows for inter-orbital kinetic
coupling. In that sense, the orbital-selective Mott transition in $d$ electron
systems shares one important aspect with what happens in the heavy fermion
compounds: The realization of the orbital-selective Mott phase requires that the
electron correlations suppress the inter-orbital hybridization. This has been
stressed in the context of iron-based superconductors \cite{Yu17.1}.

\subsection{``Flat'' bands and electron correlations}\label{implicationsB}
 
In heavy fermion systems, the $f$ electrons experience a large local Coulomb
interaction $U$ (including the Hubbard and Hund's coupling) and, in addition,
their bare bandwidth $W_f$ is small. The enhanced $U/W_f$ is responsible for the
strongly correlated nature of the $f$ electron system.

In $d$ electron systems, one may be able to use the crystalline lattice geometry
to produce narrow bands and, thus, to engineer enhanced correlation effect. In
recent years, this direction has been explored in a number of materials,
including several Fe-based kagome-lattice materials \cite{Kan20.2,Yao18.1x}.
While the focus has been on the identification of Weyl nodes in their
magnetically ordered states, one can envision the ``flat" bands to provide a
setting for strong correlation physics that may bear some analogy with what has
been so extensively studied in heavy fermion metals.

Beyond the standard correlated electron cases, twisted bilayer graphene has
provided a synthetic setting for narrow bands and strong correlation physics. 
When two layers of graphene are twisted to special angles \cite{Bis11.1},
moir\'{e} bands are formed with a small bandwidth $W_{\rm m}$. Even though the
$p$ electrons have relatively small Coulomb repulsion, which is smaller still
due to the spatially extended nature of the electronic states in the moir\'{e}
bands, $U/W_{\rm m}$ can be drastically enhanced compared to their bare graphene
counterpart. In practice, $U/W_{\rm m}$ in the ``magic angle'' twisted bilayer
graphene is of order unity. The emergence of a dome of superconductivity 
\cite{Cao18.1,Yan19.1,Lu19.2}, nematic correlations
\cite{Ker19.1,Cho19.1,Che20.1x}, and other phenomena establish these systems as
a new play ground to explore strong correlation physics. The intertwining of
correlations and topology adds a further layer of richness to the physics of
these systems.

\subsection{Strong electron correlations boosted by other
interactions}\label{implicationsC}

As discussed earlier for the heavy fermion case, strong correlations create
sensitivity also to other interactions. For instance, phonon-boosting effects
have been suggested for a number of unconventional superconductors. In FeSe, a
dramatic enhancement of the superconducting gap-opening temperature---from 8\,K
in bulk FeSe to nearly 70\,K when a single-unit-cell FeSe layer is grown on
SrTiO$_3$ \cite{Wan12.2}---was attributed in part to the coupling of electrons
in FeSe to a phonon mode in SrTiO$_3$. This proposal has received support by the
appearance of strong shaddow bands (Fig.\,\ref{hybrid}d) \cite{Lee14.1}, and has
triggered further work; for instance, recent theoretical work has suggested that
the small-momentum electron-phonon interaction underlying the replica bands
boosts but cannot trigger superconductivity in this system \cite{Li19.4}.
Whether phonons also boost superconductivity in other 2D systems is a topic of
active debate \cite{Pol19.1}.

Also photons may enhance electron interactions, an effect investigated using
ultrafast pump--probe experiments with intense light pulses in different
frequency ranges \cite{Gia16.1}. Such experiments have demonstrated transitions
between (transient) phases, for instance insulator--metal or
metal-superconductor transitions. Pioneering work on the non-superconducting
stripe-ordered cuprate La$_{1.675}$Eu$_{0.2}$Sr$_{0.125}$CuO$_4$ suggested
transiently enhanced interlayer tunneling only 1-2\,ps after irradiation with an
intense mid-infrared femtosecond pulse \cite{Fau11.1}. Very recently, evidence
for metastable light-induced superconductivity has been put forward, again in a
stripe-ordered cuprate La$_{1.885}$Ba$_{0.115}$CuO$_4$ (Fig.\,\ref{hybrid}f)
\cite{Cre19.1} but also in K$_3$C$_{60}$ \cite{Bud20.1x}. In the former, the
effect was attributed to the melting of the stripe order, delocalizing halted
electrons that then form Cooper pairs. In the latter, where the pump photon
energy was at least one order of magnitude larger than the low temperature
equilibrium superconducting gap, coupling of light with high-energy excitations,
either molecular vibrations or collective electronic modes, was suggested.

\subsection{Strong correlations and topological states}\label{implicationsD}

We have highlighted how heavy fermion metals display a rich variety of quantum
phases. A tip of iceberg is now seen in metallic phases that are topological,
particularly the Weyl-Kondo semimetal. The appearance of topological semimetals 
raises an intriguing question---how does electronic topology enrich strongly
correlated electron physics? For example, do strongly correlated topological
semimetals lead to new varieties of superconductors?

Conversely, the development of the Weyl-Kondo semimetal phase illustrates how
strong correlations drive topological states. In realizing this phase, the Kondo
effect cooperates not only with the spin-orbit coupling but also with the
nonsymmorphic space group. This points to a general strategy, namely that strong
correlations cooperate with space group symmetry in producing correlated
topological states of matter. We expect much exploration to take place on this
strategy in heavy fermion metals as well as in a variety of other strongly
correlated electron systems.

\subsection{Wider contexts}\label{implicationsE}
 
Strongly correlated systems display amplified responses to external stimuli. A
defining property of heavy fermion metals is the large effective mass of their
charge carriers, which is typically enhanced from the noninteracting value by
three orders of magnitude. The amplification occurs in response to external
parameters such as magnetic field, pressure, or doping. For Weyl-Kondo
semimetals, we have mentioned the giant responses of both thermodynamic and
magnetotransport properties. This illustrates the extreme sensitivity that
characterizes strongly correlated states of matter, which may be important for
applications in quantum technology.

Finally, it is instructive to place the study of strongly correlated systems in
the overall context of research in quantum materials and beyond. In quantum
materials we let nature work for us and reveal new physics that may not have
been conceived by our imagination. The corresponding models and concepts could
then be studied by, e.g., cold atom \cite{Bro19.1}, photonic \cite{Oza19.1}, or
gravity systems. For example, cold atom systems can be used to emulate bad
metals and strange metals (Fig.\,\ref{outreach}f). The emulated states can then
be used for measurements that are challenging to perform in quantum materials. 
A case in point are quantum entanglement properties, which are easier to measure
in cold atom systems \cite{Isl15.1}.

%%%%%%%%%%%%%%%%%%%%%%%%%%%%%%%%%%%%%%%%%%%%%%%%%%%%%%%%%%%%%%%%%%%%%%%%%%%%%%
%%%%%%%%%%%% SECTION VII %%%%%%%%%%%%%%%%%%%%%%%%%%%%%%%%%%%%%%%%%%%%%%%%%%%%%
%%%%%%%%%%%%%%%%%%%%%%%%%%%%%%%%%%%%%%%%%%%%%%%%%%%%%%%%%%%%%%%%%%%%%%%%%%%%%%

\section{Summary and outlook}\label{summary}

We have surveyed the considerable recent developments on quantum phases and
fluctuations produced by strong correlations, using heavy fermion metals as a
platform. The focus has been on the particularly interesting situation where
quantum criticality---frequently seen in these systems at the border of
antiferromagnetic order---involves quantum fluctuations that go beyond the
suppression of the Landau order parameter, in the form of a destruction of the
static Kondo effect. Because the Kondo effect involves charge and spin, breaking
it up does not only create local moments that can order but also entails an
electron localization--delocalization transition. The associated quantum
critical behavior continues to reveal new surprises, including dynamical scaling
in the charge channel. Even though these systems are metallic, the presence of
local moments allows for frustration and dimensionality as new axes of tuning.
Experiments have thus probed a considerable phase space, and the understanding
of the discovered phases and fluctuations is guided by theoretical work on the
global phase diagram. There has been considerable new development in this
delineation, from the realization of a frustration-induced strange metal phase
to the sequential localization--delocalization from entwined spin and orbital
degrees of freedom.

Quantum critical fluctuations turn the electronic system soft, and may therefore
nucleate new phases, with unconventional superconductivity as prime example. 
There has been much excitement about how the electronic
localization--delocalization transition and the associated drastic Fermi surface
transformation influences unconventional superconductivity. More generally, in
the realm of emergent phases of metallic systems, recent studies seem to have
seen a tip of iceberg in how strong correlations drive topological phases.

These developments set the stage for answering a set of fascinating outstanding
questions that are pertinent to strongly correlated metals beyond the heavy
fermion settings. We close the article by listing a few of them:\\[-0.9cm]
\begin{itemize}
\item How broadly relevant is the electronic localization--delocalization
transition to strongly correlated electron systems? For instance, how to
establish that the Fermi surface transformation observed in the high-$T_c$
cuprates and other Mott-Hubbard systems reflects the amplified quantum
fluctuations associated with electrons on the verge of localization? What are
the pertinent tuning axes for an overall zero-temperature phase diagram that
delineates the concurrence or detachment between the
localization--delocalization transition and the development of electronic
orders?\\[-0.9cm]
\item We have hypothesized a strategy for strong correlations to produce
topological states. In addition to the heavy fermion semimetals, what are the
materials platforms that can be used to explore this strategy? Can an overall
phase diagram be devised that delineates strong-correlation driven topological
phases and classifies their fluctuations?\\[-0.9cm]
\item The rich phenomena observed in strongly correlated materials comes with
considerable complexity. When disentangled, it leads to new and much clearer
understanding. How can synthetic platforms such as cold atom or mesoscopic
systems be most efficiently used to elucidate the underlying
simplicity?\\[-0.9cm]
\item We have pointed out the extreme sensitivity of strongly correlated systems
to external stimuli and their giant responses,  in particular if nontrivial
topology is involved. Can this be exploited from quantum devices? 
\end{itemize}

\vspace{0.8cm}

\noindent{\bf Acknowledgments}
We would like to thank the late Elihu Abrahams, Pegor Aynajian, Peter Blaha, Ang
Cai, Piers Coleman, Jianhui Dai, Wenxin Ding, Sami Dzsaber, Gaku Eguchi, Sven
Friedemann, Pallab Goswami, Sarah Grefe, Kai Grube, Karsten Held, Haoyu Hu, 
Kevin Ingersent, Stefan Kirchner, Jun Kono, Hsin-Hua Lai, Chia-Chuan Liu, 
Yongkang Luo, Valentina Martelli, Emilia Morosan, Andriy Nevidomskyy, Duy Ha
Nguyen, Emil Nica, Jed Pixley, Lukas Prochaska, Andrey Prokofiev, Erwin
Schuberth, Andrea Severing, Toni Shiroka, Frank Steglich, Oliver Stockert,
Liling Sun, Peijie Sun, Mathieu Taupin, Joe D. Thompson, Jan Tomczak, Hilbert
von L\"ohneysen, Steffen Wirth, Jianda Wu, Zhuan Xu, Xinlin Yan, Rong Yu, Huiqiu
Yuan, Jian-Xin Zhu, and Diego Zocco for collaborations and/or discussions. The
work has been supported in part by the Austrian Science Fund grants No.\
P29279-N27, P29296-N27, and DK W1243 and the European Union's Horizon 2020
Research and Innovation Programme Grant EMP-824109 (S.P.), and by the NSF Grant
No.\ DMR-1920740 and the Robert A.\ Welch Foundation Grant No.\ C-1411 (Q.S.).
We acknowledge the hospitality of the Aspen Center for Physics, which is
supported by the NSF grant No.~PHY-1607611.

\newpage

%%%%%%%%%%%%%%%%%%%%%%%%%%%%%%%%%%%%%%%%%%%%%%%%%%%%%%%%%%%%%%%%%%%%%%%%%%%%%%

%\bibliographystyle{naturemagallauthors}
%\bibliography{../silke_all}

\clearpage

\newpage

%%%%%%%%%%%%%%%%%%%%%%%%%%%%%%%%%%%%%%%%%%%%%%%%%%%%%%%%%%%%%%%%%%%%%%%%%%%%%%
%%%%%%%%%%%% FIGURE 1: materials %%%%%%%%%%%%%%%%%%%%%%%%%%%%%%%%%%%%%%%%%%%%%
%%%%%%%%%%%%%%%%%%%%%%%%%%%%%%%%%%%%%%%%%%%%%%%%%%%%%%%%%%%%%%%%%%%%%%%%%%%%%%
%%%%%%%%%%%%%%%%%%%%%%%%%%%%%%%%%%%%%%%%%%%%%%%%%%%%%%%%%%%%%%%%%%%%%%%%%%%%%%
%%%%%%%%%%%%%%%%%%%%%%%%%%%%%%%%%%%%%%%%%%%%%%%%%%%%%%%%%%%%%%%%%%%%%%%%%%%%%%
\clearpage
\newpage

%%%%%%%%%%%%%%%%%%%%%%%%%%%%%%%%%%%%%%%%%%%%%%%%%%%%%%%%%%%%%%%%%%%%%%%%%%%%%%
%%%%%%%%%%%% FIGURE 2: tuning %%%%%%%%%%%%%%%%%%%%%%%%%%%%%%%%%%%%%%%%%%%%%%%%
%%%%%%%%%%%%%%%%%%%%%%%%%%%%%%%%%%%%%%%%%%%%%%%%%%%%%%%%%%%%%%%%%%%%%%%%%%%%%%
\clearpage
\newpage

\begin{center}
cont. FIG.\ 2: ... and graphene. {\bf e} Correlation strength in various
topological materials as quantified by the reduced Drude spectral weight
compared to the DFT value. The symbol for twisted bilayer graphene (BLG)
represents a prediction from the observed insulating state. References for all
materials in {\bf d} and {\bf e} other then ZrSiSe are given in
ref.\citenum{Sha20.1} (adapted with permission from ref.\citenum{Sha20.1}).
\end{center}
%%%%%%%%%%%%%%%%%%%%%%%%%%%%%%%%%%%%%%%%%%%%%%%%%%%%%%%%%%%%%%%%%%%%%%%%%%%%%%
%%%%%%%%%%%%%%%%%%%%%%%%%%%%%%%%%%%%%%%%%%%%%%%%%%%%%%%%%%%%%%%%%%%%%%%%%%%%%%
\clearpage
\newpage

%%%%%%%%%%%%%%%%%%%%%%%%%%%%%%%%%%%%%%%%%%%%%%%%%%%%%%%%%%%%%%%%%%%%%%%%%%%%%%
%%%%%%%%%%%% FIGURE 3: entropy %%%%%%%%%%%%%%%%%%%%%%%%%%%%%%%%%%%%%%%%%%%%%%%
%%%%%%%%%%%%%%%%%%%%%%%%%%%%%%%%%%%%%%%%%%%%%%%%%%%%%%%%%%%%%%%%%%%%%%%%%%%%%%
%%%%%%%%%%%%%%%%%%%%%%%%%%%%%%%%%%%%%%%%%%%%%%%%%%%%%%%%%%%%%%%%%%%%%%%%%%%%%%
%%%%%%%%%%%%%%%%%%%%%%%%%%%%%%%%%%%%%%%%%%%%%%%%%%%%%%%%%%%%%%%%%%%%%%%%%%%%%%
\clearpage
\newpage

%%%%%%%%%%%%%%%%%%%%%%%%%%%%%%%%%%%%%%%%%%%%%%%%%%%%%%%%%%%%%%%%%%%%%%%%%%%%%%
%%%%%%%%%%%% FIGURE 4: FSjump %%%%%%%%%%%%%%%%%%%%%%%%%%%%%%%%%%%%%%%%%%%%%%%%
%%%%%%%%%%%%%%%%%%%%%%%%%%%%%%%%%%%%%%%%%%%%%%%%%%%%%%%%%%%%%%%%%%%%%%%%%%%%%%
\begin{figure}[h!]
\centering
\begin{minipage}[t]{0.66\textwidth}\raggedright
\vspace{0cm}

\includegraphics[width=1.0\textwidth]{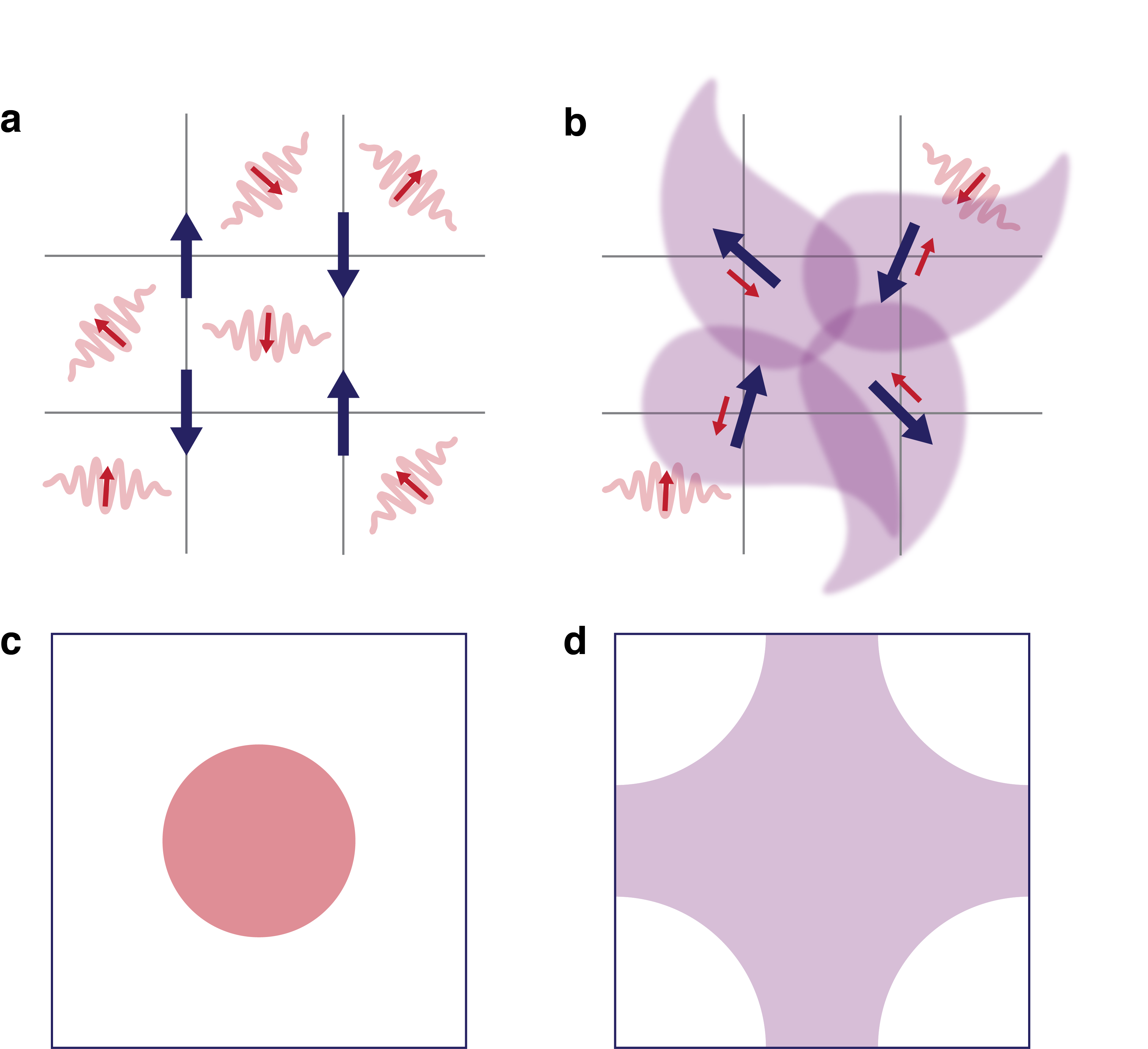}
\vspace{-0.2cm}

\hspace{0.4cm}\includegraphics[width=0.86\textwidth]{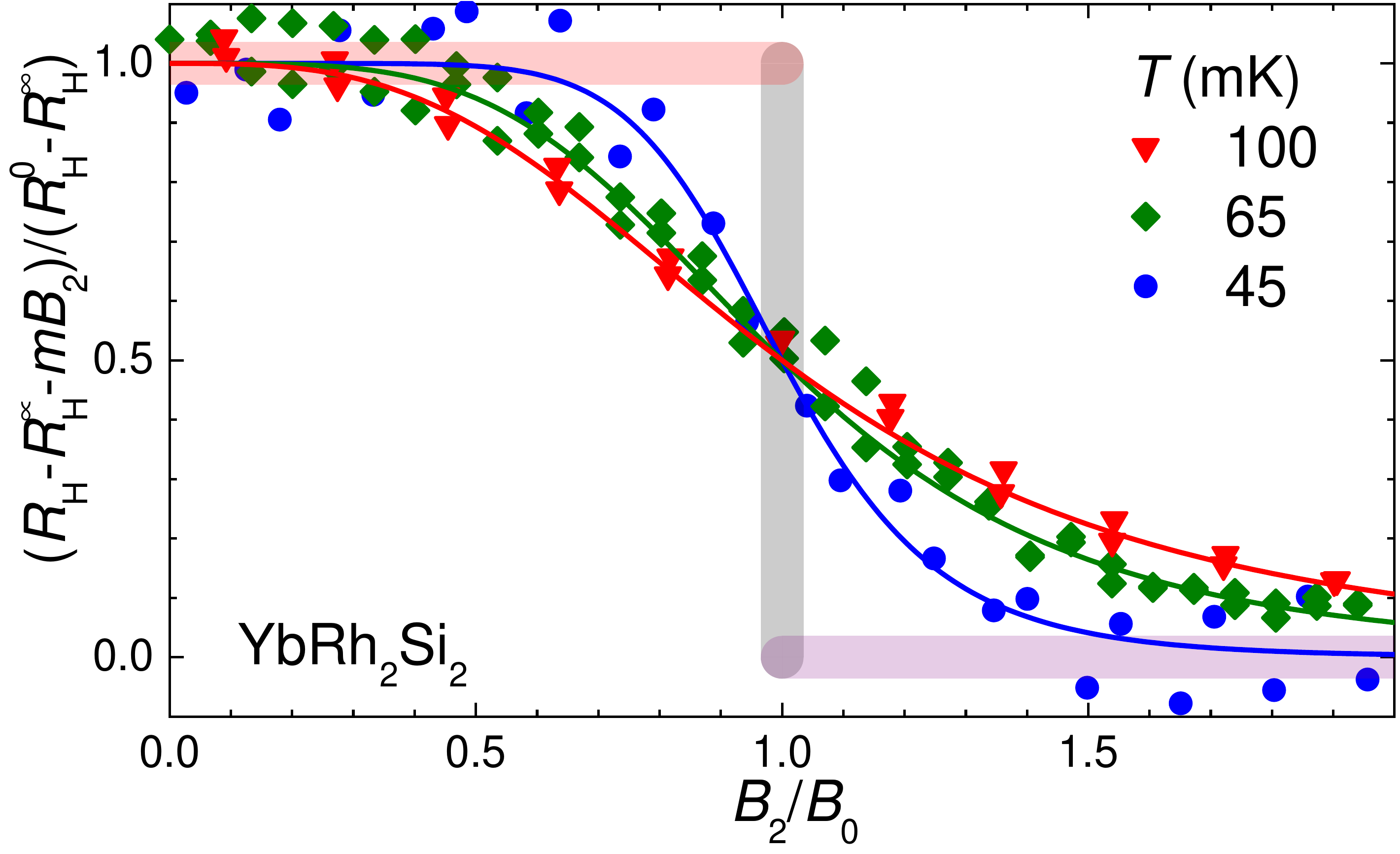}
\vspace{-6.5cm}

\hspace{0cm}{\bf{\textsf{e}}}
\end{minipage}\hfill
\begin{minipage}[t]{0.33\textwidth}\raggedright
\vspace{1.23cm}

\centerline{\hspace{-0.5cm}\includegraphics[width=0.956\textwidth]{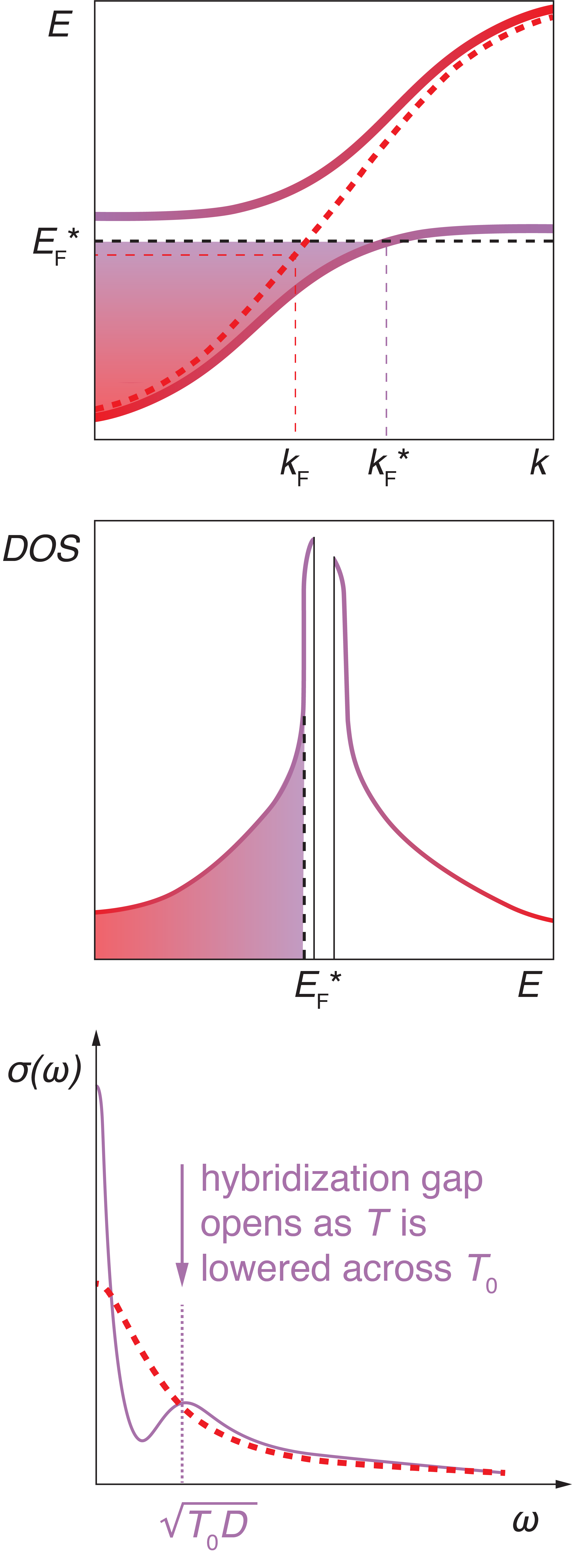}}
\vspace{-14.95cm}

\hspace{-0.4cm}{\bf\textsf{f}}
\vspace{4.13cm}

\hspace{-0.4cm}{\bf\textsf{g}}
\vspace{3.8cm}

\hspace{-0.4cm}{\bf\textsf{h}}
\end{minipage}
\vspace{5.6cm}

\caption{\label{FSjump} {\bf Fermi-surface jump at a Kondo destruction QCP.} 
{\bf a-b}$|$ Kondo lattice model, in its Kondo-destroyed and Kondo-screened
phases, respectively. The dark blue arrows label the local spins, and the red
arrows with wavy lines represent conduction electrons. {\bf c-d}$|$ Small and
large Fermi surfaces, respectively. {\bf e}$|$ The linear-response isothermal
Hall coefficient as function of a tuning magnetic field $B_2$. The latter is
normalized by the threshold field for the QCP, $B_0$. The jump is the limit of
extrapolating the data to the $T=0$ limit (adapted with permission from
ref.\citenum{Fri10.2}). {\bf f-h}$|$ Kondo screened state with hybridized bands,
a close-up of the corresponding density of states near the gap, and its optical
conductivity signature, respectively. Panel {\bf h} is adapted with permission
from ref.\citenum{Kir20.1}.}
\end{figure}
%%%%%%%%%%%%%%%%%%%%%%%%%%%%%%%%%%%%%%%%%%%%%%%%%%%%%%%%%%%%%%%%%%%%%%%%%%%%%%
%%%%%%%%%%%%%%%%%%%%%%%%%%%%%%%%%%%%%%%%%%%%%%%%%%%%%%%%%%%%%%%%%%%%%%%%%%%%%%
\clearpage
\newpage

%%%%%%%%%%%%%%%%%%%%%%%%%%%%%%%%%%%%%%%%%%%%%%%%%%%%%%%%%%%%%%%%%%%%%%%%%%%%%%
%%%%%%%%%%%% FIGURE 5: scaling %%%%%%%%%%%%%%%%%%%%%%%%%%%%%%%%%%%%%%%%%%%%%%%
%%%%%%%%%%%%%%%%%%%%%%%%%%%%%%%%%%%%%%%%%%%%%%%%%%%%%%%%%%%%%%%%%%%%%%%%%%%%%%
%%%%%%%%%%%%%%%%%%%%%%%%%%%%%%%%%%%%%%%%%%%%%%%%%%%%%%%%%%%%%%%%%%%%%%%%%%%%%%%
%%%%%%%%%%%%%%%%%%%%%%%%%%%%%%%%%%%%%%%%%%%%%%%%%%%%%%%%%%%%%%%%%%%%%%%%%%%%%%
\clearpage
\newpage

%%%%%%%%%%%%%%%%%%%%%%%%%%%%%%%%%%%%%%%%%%%%%%%%%%%%%%%%%%%%%%%%%%%%%%%%%%%%%%
%%%%%%%%%%%% FIGURE 6: hybrid %%%%%%%%%%%%%%%%%%%%%%%%%%%%%%%%%%%%%%%%%%%%%%%%
%%%%%%%%%%%%%%%%%%%%%%%%%%%%%%%%%%%%%%%%%%%%%%%%%%%%%%%%%%%%%%%%%%%%%%%%%%%%%%
%%%%%%%%%%%%%%%%%%%%%%%%%%%%%%%%%%%%%%%%%%%%%%%%%%%%%%%%%%%%%%%%%%%%%%%%%%%%%%
%%%%%%%%%%%%%%%%%%%%%%%%%%%%%%%%%%%%%%%%%%%%%%%%%%%%%%%%%%%%%%%%%%%%%%%%%%%%%%
\clearpage
\newpage

\begin{center}
cont. FIG.\ 6: ...  (full lines, adapted with permission from
ref.\citenum{Lee14.1}). {\bf e}$|$ Thermopower at 300\,K of the type-I clathrate
Ce$_{1.1}$Ba$_{6.9}$Au$_{5.5}$Si$_{40.5}$ (red symbol) compared with non-$4f$
reference compounds (black, violet) of various charge carrier concentrations $n$
(adapted with permission from ref.\citenum{Pro13.1}). {\bf f}$|$ Superfluid
density vs product of zero-frequency conductivity and superconducting transition
temperature, including what is interpreted as the transition temperature after pumping, for several dopings of La$_{2-x}$Ba$_x$CuO$_4$, providing evidence for light-boosted superconductivity (reproduced with permission from
ref.\citenum{Cre19.1}).
\end{center}

\clearpage
\newpage

%%%%%%%%%%%%%%%%%%%%%%%%%%%%%%%%%%%%%%%%%%%%%%%%%%%%%%%%%%%%%%%%%%%%%%%%%%%%%%
%%%%%%%%%%%% FIGURE 7: WeylKondo %%%%%%%%%%%%%%%%%%%%%%%%%%%%%%%%%%%%%%%%%%%%%
%%%%%%%%%%%%%%%%%%%%%%%%%%%%%%%%%%%%%%%%%%%%%%%%%%%%%%%%%%%%%%%%%%%%%%%%%%%%%%
%%%%%%%%%%%%%%%%%%%%%%%%%%%%%%%%%%%%%%%%%%%%%%%%%%%%%%%%%%%%%%%%%%%%%%%%%%%%%%
%%%%%%%%%%%%%%%%%%%%%%%%%%%%%%%%%%%%%%%%%%%%%%%%%%%%%%%%%%%%%%%%%%%%%%%%%%%%%%
\clearpage
\newpage

%%%%%%%%%%%%%%%%%%%%%%%%%%%%%%%%%%%%%%%%%%%%%%%%%%%%%%%%%%%%%%%%%%%%%%%%%%%%%%
%%%%%%%%%%%% FIGURE 8: outreach %%%%%%%%%%%%%%%%%%%%%%%%%%%%%%%%%%%%%%%%%%%%%%
%%%%%%%%%%%%%%%%%%%%%%%%%%%%%%%%%%%%%%%%%%%%%%%%%%%%%%%%%%%%%%%%%%%%%%%%%%%%%%
\clearpage
\newpage

\begin{center}
cont. FIG.\ 8: ... YBCO (grey and red squares), and strongly overdoped Tl-2201
(white diamond, adapted with permission from ref.\citenum{Bad16.1}). {\bf d}$|$
Inverse effective mass (yellow and white symbols, from quantum oscillations) and
$T_{\rm c}$ (blue circles, from resistivity) vs hole doping in
YBa$_2$Cu$_3$O$_{6+\delta}$ (reproduced with permission from
ref.\citenum{Ram15.1}). {\bf e}$|$ Temperature-pressure phase diagram of
NaYbSe$_2$ with regions featuring quantum spin liquid (QSL), weak localization
(WL), non-Fermi liquid (NFL), superconducting (SC), and Fermi liquid (FL)
behavior (reproduced with permission from ref.\citenum{Jia20.1}). {\bf f}$|$
Resistivity vs temperature of ultracold lithium-6 atoms in a two-dimensional
optical lattice, from experiment (red points), finite-temperature Lanczos
calculations (blue), and single-site DMFT (green); the gray line represents the
upper bound on Drude resistivity (reproduced with permission from
ref.\citenum{Bro19.1}).
\end{center}
%%%%%%%%%%%%%%%%%%%%%%%%%%%%%%%%%%%%%%%%%%%%%%%%%%%%%%%%%%%%%%%%%%%%%%%%%%%%%%
%%%%%%%%%%%%%%%%%%%%%%%%%%%%%%%%%%%%%%%%%%%%%%%%%%%%%%%%%%%%%%%%%%%%%%%%%%%%%%
\clearpage
\newpage

%%%%%%%%%%%%%%%%%%%%%%%%%%%%%%%%%%%%%%%%%%%%%%%%%%%%%%%%%%%%%%%%%%%%%%%%%%%%%%
%%%%%%%%%%%% SUPPLEMENTARY INFORMATION %%%%%%%%%%%%%%%%%%%%%%%%%%%%%%%%%%%%%%%
%%%%%%%%%%%%%%%%%%%%%%%%%%%%%%%%%%%%%%%%%%%%%%%%%%%%%%%%%%%%%%%%%%%%%%%%%%%%%%

{\LARGE{Supplementary Information}}\label{SI}

\vspace{0.5cm}

%%%%%%%%%%%%%%%%%%%%%%%%%%%%%%%%%%%%%%%%%%%%%%%%%%%%%%%%%%%%%%%%%%%%%%%%%%%%%%
%%%%%%%%%%%% FIGURE S1: phase diagrams %%%%%%%%%%%%%%%%%%%%%%%%%%%%%%%%%%%%%%%
%%%%%%%%%%%%%%%%%%%%%%%%%%%%%%%%%%%%%%%%%%%%%%%%%%%%%%%%%%%%%%%%%%%%%%%%%%%%%%
\begin{figure}[h!]
\centering
\centerline{\includegraphics[height=0.26\textheight]{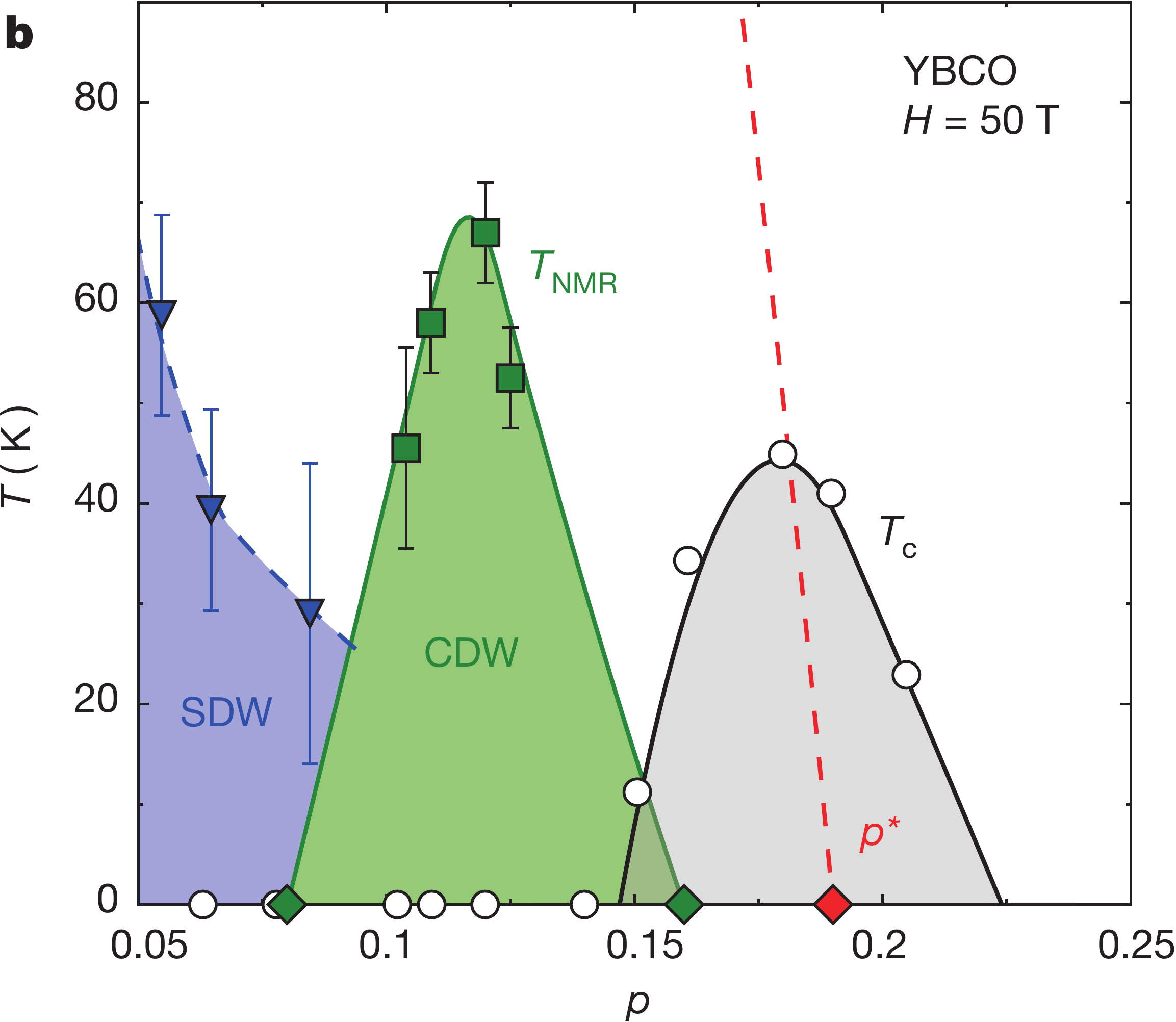}\hspace{0.5cm}\includegraphics[height=0.265\textheight]{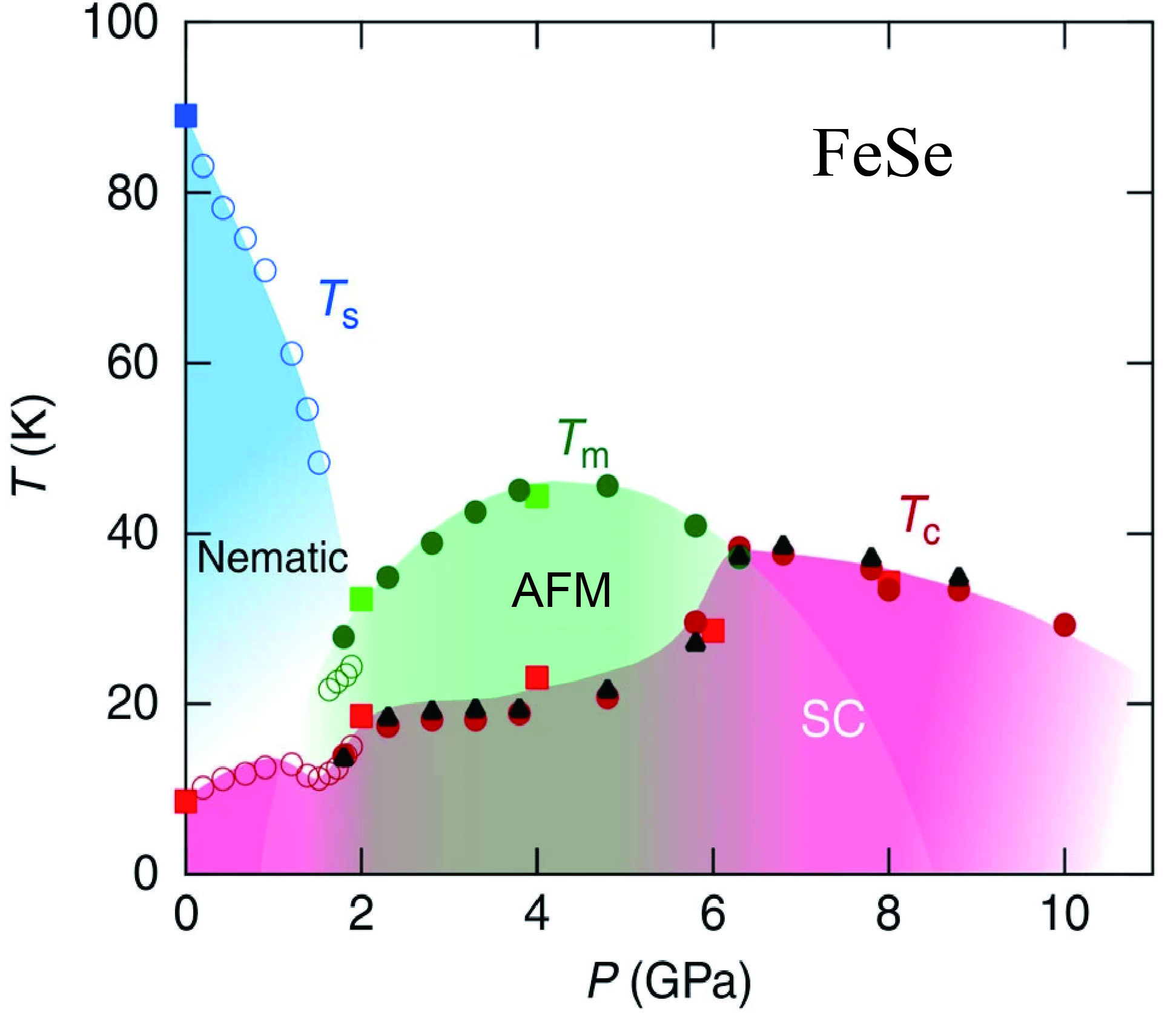}}
\vspace{0.2cm}

\centerline{\includegraphics[height=0.25\textheight]{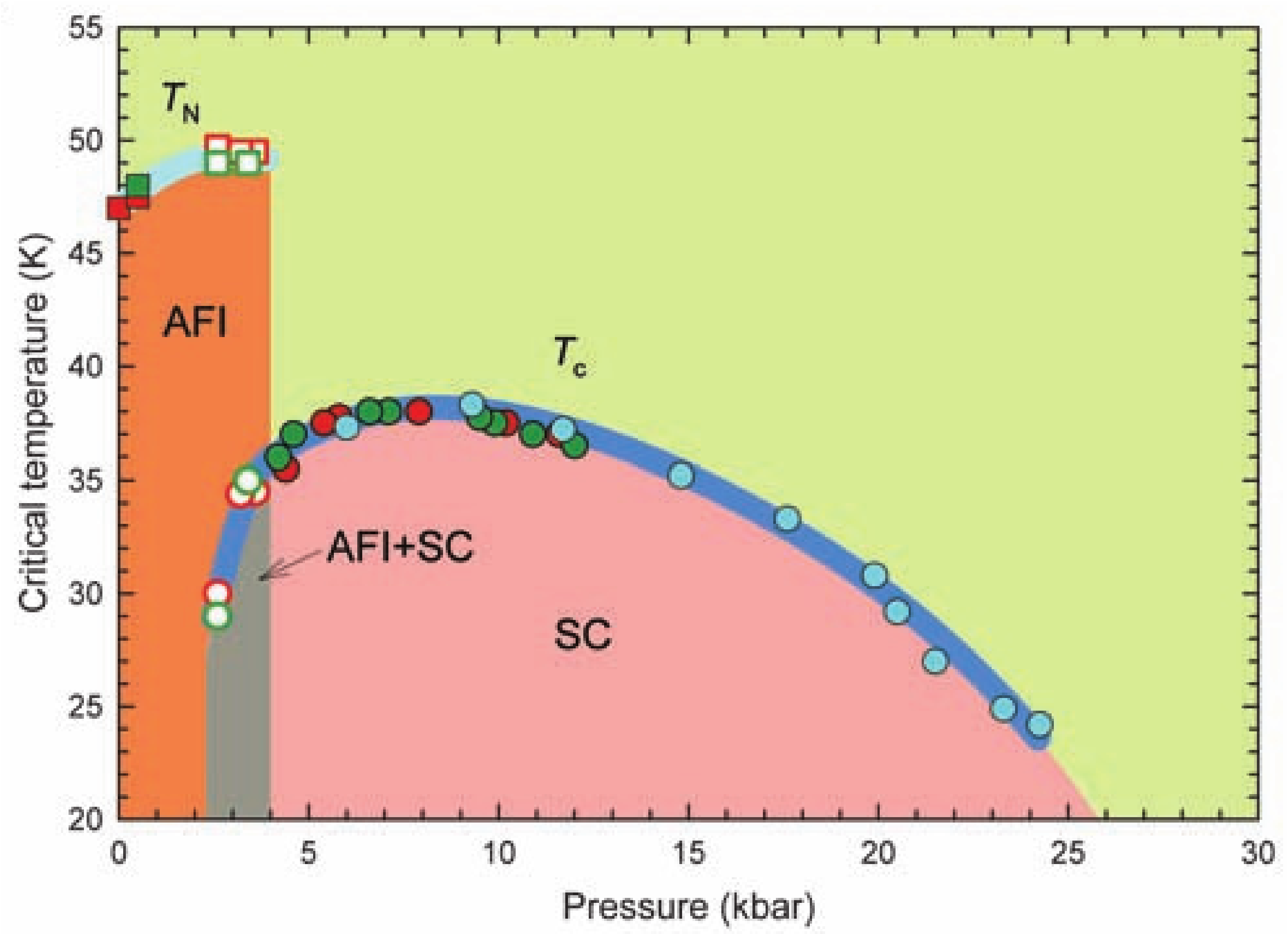}\hspace{0.5cm}\includegraphics[height=0.27\textheight]{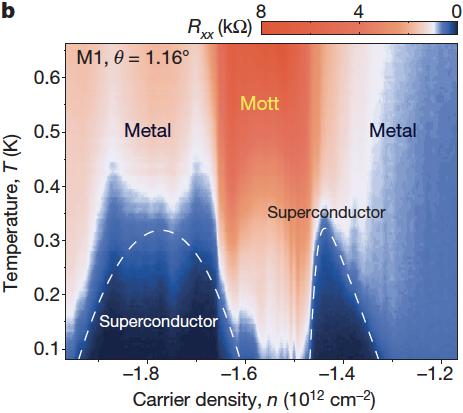}}
\vspace{-13.3cm}

\hspace{-14.3cm}\includegraphics*[width=0.06\textwidth]{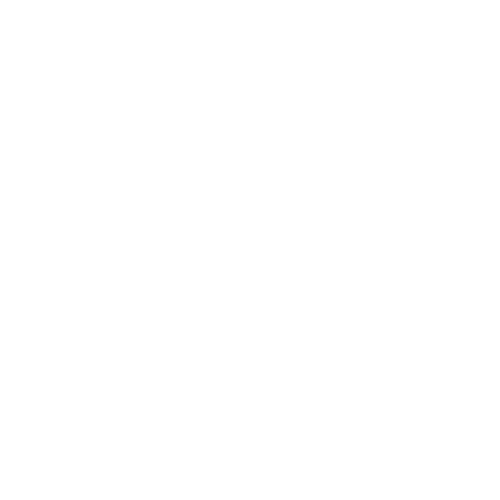}
\vspace{-0.9cm}

\hspace{-7.2cm}{\bf\textsf{a}}\hspace{7.4cm}{\bf\textsf{b}}
\vspace{6.cm}

\hspace{2cm}\includegraphics*[width=0.06\textwidth]{whitebox.pdf}
\vspace{-1cm}

\hspace{-7cm}{\bf\textsf{c}}\hspace{8.3cm}{\bf\textsf{d}}
\vspace{5.7cm}

\end{figure}
FIG.\,S1: {\bf Landscape of quantum phases in strongly correlated materials.}
{\bf a-d}$|$ Temperature-tuning parameter phase diagrams of the cuprate
superconductor YBaCuO$_{6-\delta}$ (reproduced with permission from
ref.\citenum{Bad16.1}), the iron pnictide FeSe (adapted with permission from
ref.\citenum{Sun16.1}), the fulledide compound C$_{60}$ (reproduced with
permission from ref.\citenum{Tak09.1}), and magic angle twisted bilayer graphene
(reproduced with permission from ref.\citenum{Cao18.1}), respectively. The
different phases and functionalities result from the interplay of various
low-energy degrees of freedom as illustrated in Fig.\,1 of the main part.
%%%%%%%%%%%%%%%%%%%%%%%%%%%%%%%%%%%%%%%%%%%%%%%%%%%%%%%%%%%%%%%%%%%%%%%%%%%%%%
%%%%%%%%%%%%%%%%%%%%%%%%%%%%%%%%%%%%%%%%%%%%%%%%%%%%%%%%%%%%%%%%%%%%%%%%%%%%%%

%%%%%%%%%%%%%%%%%%%%%%%%%%%%%%%%%%%%%%%%%%%%%%%%%%%%%%%%%%%%%%%%%%%%%%%%%%%%%%
%%%%%%%%%%%% BOX 1 - part 1 %%%%%%%%%%%%%%%%%%%%%%%%%%%%%%%%%%%%%%%%%%%%%%%%%%
%%%%%%%%%%%%%%%%%%%%%%%%%%%%%%%%%%%%%%%%%%%%%%%%%%%%%%%%%%%%%%%%%%%%%%%%%%%%%%
\begin{InfoBox}[h]
\caption{{\bf Heavy fermion state out of Kondo lattice}\label{box_Kondo}} 
\fbox{
\begin{minipage}{0.98\textwidth}\raggedright
Heavy fermion materials contain elements with partially-filled $f$ electrons,
which are coupled to $spd$ electrons through a hybridization matrix 
\cite{Hew97.1}. In many cases, the count of the $f$ electrons is fixed at an
odd-integer value, because a change of their valence occupation costs too much
Coulomb repulsion. For example, there would be one 4$f$ electron per Ce ion or
thirteen 4$f$ electrons (which is equivalent to one 4$f$ hole) per Yb ion. In
such cases, an $f$ electron acts as a localized magnetic moment in the
low-energy description. We illustrate the physics in the simplest situation,
with one spin-$1/2$ local moment occupying each site, each unit containing only
one site, and only one $spd$ based conduction electron band. This is the Kondo
lattice Hamiltonian, which reads
\begin{eqnarray}\label{eq:KLHamiltonian}
{\cal H}_{\rm KL} 
&=& {\cal H}_f + {\cal H}_K + {\cal H}_c , \nonumber\\
{\cal H}_f &=& (1/2) \sum_{ ij} I_{ij} ~{\bf S}_{i} ~ {\bf S}_{j}\,; 
~~~~~~ {\cal H}_K = \sum_i J_{\rm K} ~{\bf S}_{i} \cdot {\bf s}_{c,i}\,;
~~~~~~ {\cal H}_c = \sum_{\bf k \sigma} \epsilon_{\bf k}
c_{{\bf k}\sigma}^{\dagger} c_{{\bf k}\sigma}
\end{eqnarray}
At a site $i$, the local moment, ${\bf S}_i$ interacts with the spin of the
conduction electron, ${\bf s}_{c,i} = (1/2) c_{i}^{\dagger} \vec{\sigma} c_i$,
through an antiferromagnetic Kondo exchange coupling $J_{\rm K}$. It is
convenient to introduce an explicit term, $I_{ij}$, to describe the RKKY
interaction between the local moments located at different sites. Finally,
$\epsilon_{\bf k}$ describes the band dispersion of the conduction $c$
electrons. The model can be generalized to incorporate additional complexity.

For the Hamiltonian \ref{eq:KLHamiltonian}, the Kondo coupling can be
parameterized by a bare Kondo energy scale, $T_0 = N_{\rm F}^{-1}
\exp{(-1/J_{\rm K} N_{\rm F})} $ (with $k_{\rm B}$ being set to unity), where 
$N_{\rm F}$ is the density of states of the noninteracting conduction electron
band at the Fermi energy [the corresponding Fermi momentum being marked $k_{\rm
F}$ in (Fig.\,\ref{FSjump}f)]. The RKKY interaction $I_{ij}$ is taken to have a
characteristic value $I$. The Doniach competition \cite{Don77.1} is
characterized by tuning the parameter $\delta = T_0/I$. Where $T_0$ dominates
over $I$ is the regime for the traditional description of a paramagnetic heavy
Fermi liquid. In the ground state, the local moments and the spins of conduction
electrons form a Kondo singlet. To describe the excitation spectrum associated
with this ground state, it is convenient to rewrite the Kondo interaction at any
given site to be (1/2)$\sum_{\sigma} [\sigma S^z c^{\dagger}_{\sigma} +
X^{\sigma\bar{\sigma}} c^{\dagger}_{\bar{\sigma}} ]c_{\sigma}$, where
$\bar{\sigma} \equiv -\sigma$, and $X^{\sigma\bar{\sigma}} $ is the spin raising
(lowering) operator for $\sigma = \uparrow$ ($\downarrow$). With a nonzero
static amplitude for such a Kondo singlet, the local moment binds with the
conduction electron, and $F^{\dagger}_{\sigma} =\sigma S^z c^{\dagger}_{\sigma}
+ X^{\sigma\bar{\sigma}} c^{\dagger}_{\bar{\sigma}}$ acts as a composite
fermion. The Kondo interaction becomes a hybridization $\sum_{\sigma}
F^{\dagger}_{\sigma} c_{\sigma}$. From the perspective of the conduction
\end{minipage}
}
\end{InfoBox}
\clearpage
\newpage

%%%%%%%%%%%%%%%%%%%%%%%%%%%%%%%%%%%%%%%%%%%%%%%%%%%%%%%%%%%%%%%%%%%%%%%%%%%%%%
%%%%%%%%%%%% BOX 1 - part 2 %%%%%%%%%%%%%%%%%%%%%%%%%%%%%%%%%%%%%%%%%%%%%%%%%%
%%%%%%%%%%%%%%%%%%%%%%%%%%%%%%%%%%%%%%%%%%%%%%%%%%%%%%%%%%%%%%%%%%%%%%%%%%%%%%

\fbox{
\begin{minipage}{0.98\textwidth}
electrons, $F^{\dagger}_{\sigma}$ introduces a resonance near the Fermi energy.
Thus, $F^{\dagger}_{\sigma}$ is referred to as creating a Kondo resonance. In
the overall electronic excitations, the composite fermion hybridizes with the
conduction electron band to form the hybridized bands sketched in
Fig.\,\ref{FSjump}f \cite{Hew97.1}. The Fermi surface is formed by both the
composite fermions and the conduction electrons, and the Fermi surface ($k_{\rm
F}^*$) becomes large. The composite fermions lead to a density of states with a
sharp peak near the Fermi energy (see Fig.\,\ref{FSjump}g), which spreads over a
narrow energy range determined by the Kondo scale.

The scale $T_0$ is also reflected in the evolution of the excitation spectrum as
a function of temperature. At temperatures much larger than $T_0$, the local
moments and conduction electrons are essentially decoupled. As temperature is
lowered through $T_0$, the Kondo correlation between the two species of spins
start to set in. This leads to the initial onset of a ``hybridization gap", as
would be seen in the optical conductivity $\sigma(\omega)$ (see
Fig.\,\ref{FSjump}h). The corresponding frequency scale is $\sqrt{T_0D}$ (where
$D$ is the conduction electron bandwidth) is determined by the zero-wavevector
(direct) transfer transition between the occupied and empty states in the
hybridized bands.

\end{minipage}
}
%%%%%%%%%%%%%%%%%%%%%%%%%%%%%%%%%%%%%%%%%%%%%%%%%%%%%%%%%%%%%%%%%%%%%%%%%%%%%%
%%%%%%%%%%%%%%%%%%%%%%%%%%%%%%%%%%%%%%%%%%%%%%%%%%%%%%%%%%%%%%%%%%%%%%%%%%%%%%
\clearpage
\newpage

%%%%%%%%%%%%%%%%%%%%%%%%%%%%%%%%%%%%%%%%%%%%%%%%%%%%%%%%%%%%%%%%%%%%%%%%%%%%%%
%%%%%%%%%%%% BOX 2 %%%%%%%%%%%%%%%%%%%%%%%%%%%%%%%%%%%%%%%%%%%%%%%%%%%%%%%%%%%
%%%%%%%%%%%%%%%%%%%%%%%%%%%%%%%%%%%%%%%%%%%%%%%%%%%%%%%%%%%%%%%%%%%%%%%%%%%%%%
\begin{InfoBox}[h]
\vspace{-0.8cm}

\caption{{\bf Noninteracting electronic topology.}\label{box_topology}} 
\fbox{
\begin{minipage}{0.98\textwidth}\raggedright
Weyl semimetals are materials in which bulk energy bands touch only at discrete
points in momentum space (the Weyl nodes) and where, in the vicinity of the
nodes, the electron wave functions of these two bands can be approximated by the
Weyl equation, which has far-reaching consequences. In the pertinent lattice
theory of chirally invariant fermions with locality, there is an equal number of
production and annihilation of Weyl fermions \cite{Nie83.1}. Thus, Weyl nodes
with Weyl fermions of opposite chirality occur in pairs such that the axial
charge is conserved. The nontrivial topological nature of a Weyl semimetal
guarantees that the two nodes of such a pair are separated in momentum space.
Experimental observables include the linear bulk dispersion near the Weyl nodes,
surface Fermi arcs that connect the projection of two bulk Weyl nodes of
different chirality in the surface Brillouin zone, the chiral anomaly
manifesting in an extremely large negative longitudinal magnetoresistance, or
the monopoles and an antimonopoles of Berry flux in momentum space leading to
spontaneous or anomalous Hall responses. Space group symmetry plays an important
role in the emergence of Weyl nodes in a robust way
\cite{You12.1,Po17.1,Can18.1}.

ARPES is playing a key role in the identification of weakly-interacting
topological electronic materials. As a highly surface sensitive technique it is
particularly well suited to probe topological surface states. Indeed, an APRES
investigation of the (111) surface of the topological insulator Bi$_2$Se$_3$
revealed, with amazing clarity, a single Dirac cone in the surface electronic
band dispersion ({\bf a}, adapted with permission from ref.\citenum{Xia09.1}).
If higher energy photons are used, ARPES acquires some bulk sensitivity. It has
thus also been used to probe topological bulk bands. Examples are the
identification of a bulk Dirac cone in Na$_3$Bi ({\bf b}, adapted with
permission from ref.\citenum{Liu14.1}) and of two closely spaced Weyl cones in
TaAs ({\bf c}, adapted with permission from ref.\citenum{Xu15.2}). In the
latter, crescent-shaped surface states ({\bf d}, adapted with permission from
ref.\citenum{Xu15.2}) were assigned as being the topological Fermi arcs that
connect the surfaces projections of the bulk Weyl nodes ($+$ and $-$ in panel
{\bf c}). That there are two arcs is attributed to two pairs of bulk nodes
collapsing in the surface projection. \vspace{0.5cm}

\centerline{\hspace{0.1cm}\includegraphics[height=0.258\textwidth]{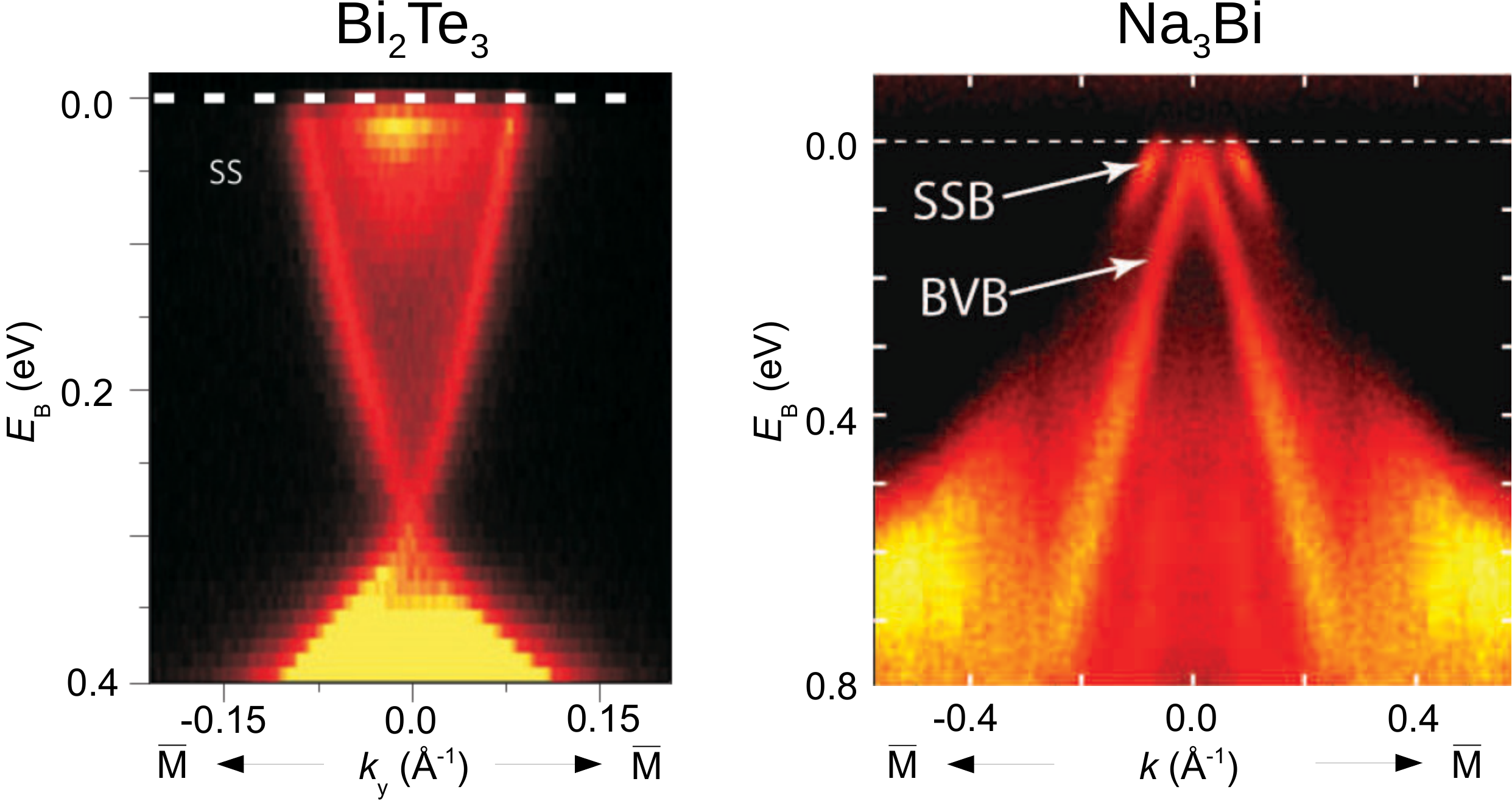}\hspace{0.5cm}\includegraphics[height=0.258\textwidth]{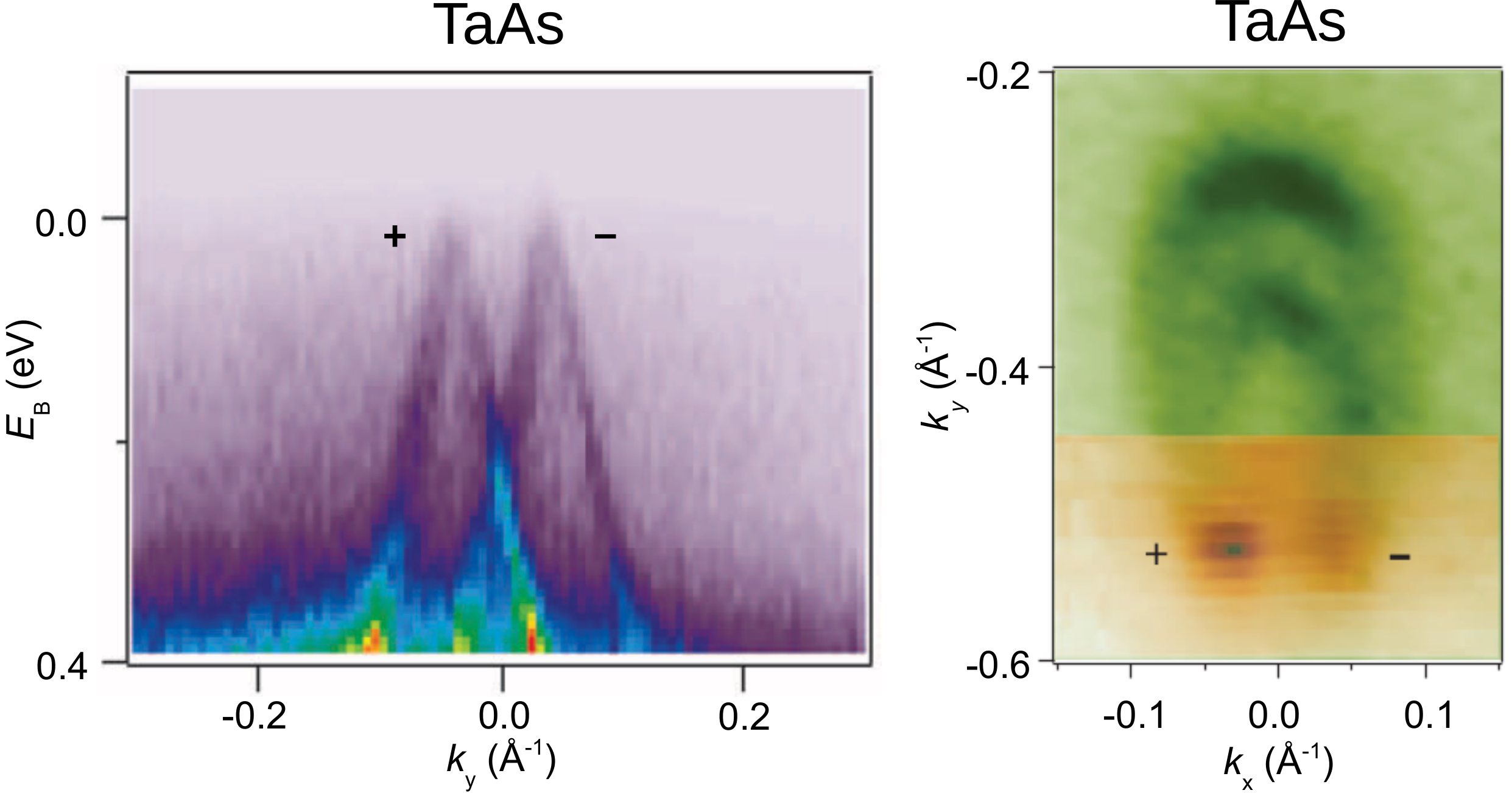}
}
\vspace{-4.7cm}

{\bf\textsf{a}}\hspace{3.6cm}{\bf\textsf{b}}\hspace{4.2cm}{\bf\textsf{c}}\hspace{4.6cm}{\bf\textsf{d}}
\vspace{4cm}

\end{minipage}
}\par
\end{InfoBox}
\newpage
%%%%%%%%%%%%%%%%%%%%%%%%%%%%%%%%%%%%%%%%%%%%%%%%%%%%%%%%%%%%%%%%%%%%%%%%%%%%%%
%%%%%%%%%%%%%%%%%%%%%%%%%%%%%%%%%%%%%%%%%%%%%%%%%%%%%%%%%%%%%%%%%%%%%%%%%%%%%%

\end{document}